\documentclass[
amssymb,preprint,preprintnumbers,nofootinbib,superscriptaddress]{revtex4}
\usepackage{amsmath}
\usepackage{graphicx}
\graphicspath{{Figures/}}
\usepackage{latexsym}
\usepackage{amsfonts}
\usepackage{url,hyperref}
\usepackage{bm}
\usepackage{textcomp}
\usepackage{color}
\usepackage{bbm}
\usepackage{slashed}
\usepackage{caption}
\usepackage{epstopdf}
\usepackage{subcaption}
\usepackage{placeins}
\usepackage{color}
\usepackage{cancel} 
\usepackage[normalem]{ulem}
\usepackage{tcolorbox}
\usepackage{soul}

\begin{document}

\title{Holographic Dark Energy from the Anti-de Sitter Black Hole}

\author{Ratchaphat Nakarachinda \footnote{Email: tahpahctar\_net@hotmail.com}}
\affiliation{The Institute for Fundamental Study, Naresuan University, Phitsanulok, 65000, Thailand}	
		
\author{Chakrit Pongkitivanichkul \footnote{Email: chakpo@kku.ac.th}}
\affiliation{Khon Kaen Particle Physics and Cosmology Theory Group (KKPaCT), Department of Physics, Faculty of Science, Khon Kaen University, 123 Mitraphap Road, Khon Kaen, 40002, Thailand}

\author{\\ Daris Samart \footnote{Email: darisa@kku.ac.th}}
\affiliation{Khon Kaen Particle Physics and Cosmology Theory Group (KKPaCT), Department of Physics, Faculty of Science, Khon Kaen University, 123 Mitraphap Road, Khon Kaen, 40002, Thailand}

\author{Lunchakorn Tannukij \footnote{Email: lunchakorn.ta@kmitl.ac.th}}
\affiliation{KOSEN-KMITL, King Mongkut Institute of Technology Ladkrabang, 1 Chalong Krung 1 Alley, Lat Krabang, Bangkok 10520, Thailand}
\affiliation{Department of Physics, School of Science, King Mongkut Institute of Technology Ladkrabang, 1 Chalong Krung 1 Alley, Lat Krabang, Bangkok 10520, Thailand}

\author{Pitayuth Wongjun \footnote{Email: pitbaa@gmail.com}}
\affiliation{The Institute for Fundamental Study, Naresuan University, Phitsanulok, 65000, Thailand}

\begin{abstract}

The anti-de Sitter (AdS) black hole plays an important role in the holographic principle. In this study, the upper bound in energy corresponding to the mass of the Schwarzschild black hole is modified to be that of the AdS black hole. Via the correspondence between the ultraviolet (UV) and infrared (IR) cutoffs, the constant term in the energy density of the holographic dark energy (HDE) can be obtained from the negative cosmological constant from the black hole. Interestingly, the proposed dark energy model could drive the late-time expansion of the Universe without the causality violation. The cosmic evolution is investigated by choosing the Hubble and particle horizons as the IR length scales. It is found that the accelerated expansion at late time can be obtained for both cases. The likelihood analysis on the model parameters is also performed. This result may shed light on the connection between the AdS black hole and the de Sitter (dS) spacetime in the context of cosmology.

\end{abstract}
\maketitle{}


\section{Introduction}\label{sec: intro}

One of the longstanding puzzles in cosmology is why the Universe expands with acceleration nowadays \cite{SupernovaSearchTeam:1998fmf,SupernovaCosmologyProject:1998vns}. Introducing a kind of exotic energy called the dark energy is a way to deal with  the accelerated expansion of the Universe based on general relativity. There are numerous attempts to construct the model of dark energy (see \cite{Li:2011sd} and references therein). One of interesting models of dark energy is based on the holographic principle, the so-called holographic dark energy (HDE). Effectively, the energy density of the HDE, $\rho$, can be described by only two physical quantities on the boundary which are the Planck mass $M_\text{P}$ and the cosmic length scale $L$. Using dimensional analysis, it can be expressed as follows:
\begin{eqnarray}
	\rho_{de}=c_1M_\text{P}^4+c_2L^{-2}M_\text{P}^2+\hdots,\label{rho de expand}
\end{eqnarray}
where $c_i$ are arbitrary dimensionless parameters. Even though the above equation is an effective equation coming from dimensional analysis, one can obtain a similar equation from the relation between UV and IR scales.

In order to see how the expression in Eq.~\eqref{rho de expand} involves the idea of the UV/IR relationship, let us consider a system in a box volume $l^3$ for a conventional quantum field theory with the UV cutoff $\Lambda_{UV}$. The entropy of this system scales with its volume, $S\sim l^3\Lambda_{UV}^3$. By the development of the black hole in the thermodynamical point of view \cite{Bekenstein:1973ur, Hawking:1975vcx, Bardeen:1973gs}, the black hole entropy has led to the upper bound of an arbitrary system with a finite size and energy \cite{Bekenstein:1980jp}. Interestingly, the volume scaling entropy of the quantum field theory is bounded by the area scaling entropy of the black hole as $l^3\Lambda_{UV}^3\lesssim S_\text{BH}\sim l^2M_\text{P}^2$ where $S_\text{BH}$ is the Bekenstein-Hawking entropy. For any value of $\Lambda_{UV}$, there exists a sufficiently large length scale compatible with the IR cutoff in which the quantum field theory breaks down \cite{tHooft:1993dmi}. It is important to note that the IR cutoff indeed depends on the UV cutoff in this aspect. There is an argument by Cohen et al. \cite{Cohen:1998zx} that the entropy of the quantum field theory never saturates the Bekenstein bound because the maximum energy of the matter in the field theory corresponds to the Schwarzschild mass, $l^3\Lambda_{UV}^4\lesssim lM_\text{P}^2$. They also proposed that the states where the energy bound is satisfied can be described by a quantum field theory \cite{Cohen:1998zx}. As a consequence of this stronger constraint, the energy density due to the vacuum fluctuation is in the order of $\Lambda_{UV}^4\lesssim l^{-2}M_\text{P}^2$.  This is actually the second term in Eq.~\eqref{rho de expand} interpreted as the energy density for the HDE,
\begin{eqnarray}
	\rho_{HDE}=3c^2L^{-2}M_\text{P}^2,\label{rho hde}
\end{eqnarray}
where $c$ is a conventional constant of the dark energy model. Moreover, the HDE can also be constructed from the covariant gravity theories \cite{Nojiri:2017opc, Nojiri:2020wmh} and the effective field theory coupled with gravity \cite{Lin:2021bxv}.

There are various kinds of HDE depending on the choices of suitable IR length scale.  By choosing the simple form of the IR length scaled as the Hubble horizon radius $H^{-1}$, the model does not admit the acceleration phase of the Universe \cite{Hsu:2004ri}. It behaves similarly as the non-relativistic matter which cannot be used to drive the accelerated expansion of the Universe.  Another class of HDE models can be constructed by choosing the length scale as the future particle horizon \cite{Li:2004rb}. The cosmological implications of this model are investigated in various aspects (see \cite{Wang:2016och} and references therein). There are a large number of studies on the models with the interaction between the HDE and dark matter as well as their observational constraints \cite{Karwan:2008ig, Ma:2009uw, Zhang:2012uu}. However, these models cannot satisfy the condition of causality, since the information of the Universe in the future is required in order to explain the evolution nowadays \cite{Colgain:2021beg}. Due to this reason, other choices of the IR length scale have become a center of attention recently, e.g., the apparent horizon for the non-flat Universe \cite{Sheykhi:2009zv}, conformal-age-like length \cite{Huang:2012nz}, total coming horizon \cite{Huang:2012xma}, and other combinations of length scales \cite{Nojiri:2019skr, Nojiri:2021iko, Nojiri:2021jxf}. HDE in various kinds of modified gravity theories are also active in research areas, e.g., the scalar-tensor theories \cite{Nojiri:2005pu, Banerjee:2007zd, Bisabr:2008gu, Kritpetch:2020vea}  and braneworld theories \cite{Saridakis:2007ns, Farajollahi:2014hzp}. Other kinds of the dark energy models are possible to construct based on the holographic principle. The models known as the agegraphic dark energy has been proposed by using time scales as the IR cutoff \cite{Cai:2007us, Wei:2007ty}. Since the Ricci scalar $R$ has indeed a dimension of $L^{-2}$, it is possible to use the IR length scale as $R^{-1/2}$. This model is the so-called Ricci dark energy model \cite{Gao:2007ep, Granda:2008dk} while the cosmic perturbations have been investigated in Ref.~\cite{Karwan:2011sh}.

One of the interesting HDE models considers the other thermodynamic descriptions of the black hole. Based on the standard black hole thermodynamics \cite{Bekenstein:1973ur, Hawking:1975vcx, Bardeen:1973gs}, the black hole mass $M$, satisfying the first law, $\text{d}M=T\text{d}S$, can be interpreted as the infinitesimal energy of HDE in a volume $\text{d}V$,
\begin{eqnarray}
	\rho_{HDE}\text{d}V\propto T_\text{H}\text{d}S_\text{BH}\sim L^{-1}\text{d}\big(L^2M_\text{P}^2\big),
\end{eqnarray}
where the temperature $T_\text{H}\sim L^{-1}$ is the Hawking temperature of the Schwarzschild black hole \cite{Hawking:1975vcx}. As a result, the above energy density is actually equivalent to Eq.~\eqref{rho hde}. The generalized versions of entropy proposed as the candidates of the black hole entropy lead to the modification of HDE, e.g., the  Tsallis entropy \cite{Tavayef:2018xwx}, R\'{e}nyi entropy \cite{Moradpour:2018ivi}, Barrow entropy \cite{Saridakis:2020zol}, Sharma-Mittal entropy \cite{SayahianJahromi:2018irq}, and recently a generalized version of the  non-extensive entropy \cite{Nojiri:2022aof}. It is very important to note that the energy density in Eq.~\eqref{rho hde} is modified due to the mentioned entropy only when the temperature is fixed as the Hawking temperature. However, the first law is not satisfied for these cases. If one considers the temperature associated with the first law $T=\partial M/\partial S$, the effect of the generalized entropy in the HDE energy density will vanish. In other words, the additional contributions due to entropy and temperature are cancelled in order to keep the first law valid.

In 1997, Maldacena proposed a conjecture that there is a remarkable duality between gravitational theory in higher dimensions and conformal field theory at a boundary \cite{Maldacena:1997re}. This leads to the so-called AdS/CFT correspondence (see \cite{Aharony:1999ti} for review). The striking advantage of the approach is that one can infer properties of a strongly interacting system which are often inaccessible by a usual perturbative regime. Although the original context of the correspondence was the string/M theory framework, it has been shown to successfully explain a wide range of physical phenomena. For instance, AdS/CFT is used to study the properties of the quark-gluon plasma and this has led to a prediction of the ratio between the shear viscosity and the density of the entropy \cite{Policastro:2001yc}. The result also gives an excellent agreement with the experimental data from RHIC \cite{Luzum:2008cw}. Moreover, the superconductivity, in condensed matter physics, is also successfully described by the AdS/CFT framework. Several interesting features of the high $T_c$ superconductor can be done via the calculations of properties of the black holes in the AdS$_4$ spacetime \cite{Hartnoll:2008vx,Hartnoll:2008kx}, such as equations of state, real time correlation function, transport properties, and so on; see reviews \cite{Herzog:2009xv,Horowitz:2010gk}. In addition, the connection between the AdS black hole and cosmology in the context of AdS/CFT correspondence is also studied \cite{Apostolopoulos:2008ru, Camilo:2016kxq, Khimphun:2020nkh}. Following the footsteps of history, it would be more interesting to extend the idea of AdS/CFT to some other connections between the AdS black hole and a physical system at the boundary.

Inspired by the remarkable correspondence, we propose that the properties of the AdS black holes could be related to dark energy by using the holographic principle. In this work, we use the same idea with the usual HDE but now we use the AdS black hole to seek a possible way to link the AdS/CFT correspondence to dark energy. In this sense, the infinitesimal energy of the HDE is the saturated energy of the AdS black hole via the first law of thermodynamics 
\begin{eqnarray}
	\rho_{HDE}\text{d}V\propto T_\text{AdS}\text{d}S_\text{BH}\sim \frac{1+\Lambda L^2}{ L}\text{d}\big(L^2M_\text{P}^2\big),
\end{eqnarray}
where $\Lambda$ is the cosmological constant which is positive for the AdS black hole. Interestingly, the cosmological constant in terms of the AdS black hole can play the role of a constant for dS spacetime in the cosmology context. This may shed light on the interplay between the AdS/CFT correspondence and dark energy in the context of cosmology. Moreover, it has been found that the first term on the right-hand side of Eq.~\eqref{rho de expand} can be obtained due to the existence of the cosmological constant from the black hole side. It has been found that this is an essential contribution to drive the accelerated expansion of the Universe at late time. By choosing the IR length scale as the Hubble radius and particle horizon, the standard evolution of the Universe including the radiation dominated epoch, matter dominated epoch and the late-time acceleration phase can be obtained. Moreover, this model does not encounter the causality problem, since it is not necessary to adopt the future particle horizons.

This work is organized as follows. The energy density of the HDE obtained from the IR cutoff due to the AdS black hole is derived via the thermodynamical approach in Sec.~\ref{sec: HDE energy density}. Sec.~\ref{sec: Hubble hor}, the exact cosmic evolution is presented for the model with the Hubble horizon as the length scale. For the particle horizon chosen as the length scale, the dynamics of the Universe is analyzed using both the dynamical system approach and numerical calculation in Sec.~\ref{sec: partical hor}. The observational constraints of the HDE models are investigated in Sec.~\ref{sec: Constraint}.
Finally, in Sec.~\ref{sec: conclu}, key results as well as interesting remarks are discussed and summarized.

\section{HDE density}\label{sec: HDE energy density} 
In this section, we will investigate how the HDE density can be obtained by using the notion of the AdS black hole. As we have mentioned, the AdS spacetime is a crucial ingredient in the holographic principle via the AdS/CFT correspondence. The AdS black hole is indeed a static and spherically symmetric solution obtained by solving the Einstein field equation with a cosmological constant. The line element can be expressed as follows:
\begin{eqnarray}
	\text{d}s^2=-f(r)\text{d}t^2 +f(r)^{-1}\text{d}r^2+r^2\text{d}\Omega^2, \quad	
	f(r)=1-\frac{2M}{r}+\frac{\Lambda}{3}r^2,
\end{eqnarray}
where $M$ and $\Lambda$ are the ADM mass of the black hole and the cosmological constant ($\Lambda>0$ for the asymptotically AdS spacetime), respectively. For the AdS black hole, there exists only one horizon denoted by $r_h$. The horizon can be obtained by solving $f(r_h)=0$. The mass of the black hole is therefore written in terms of the horizon radius as 
\begin{eqnarray}
	M=\frac{r_h}{2}\left(1+\frac{\Lambda}{3}r_h^2\right).
\end{eqnarray}
From the thermodynamic description of the black hole, the parameter $M$ can play the role of the internal energy of the system. Note that, in interpreting the cosmological constant as a thermodynamic variable namely pressure, the parameter $M$ will be the chemical enthalpy of the system. One of the important thermodynamic quantities of a black hole is an entropy. It is well known that the entropy of the black hole can be described by the Bekenstein-Hawking entropy written in terms of the surface area of the black hole horizon $A$ as $S_\text{BH}= A/4$. The black hole entropy in terms of the horizon radius is given by
\begin{eqnarray}
	S_\text{BH}=\pi r_h^2.
\end{eqnarray}
Equivalently to the first law of thermodynamics, the black hole's mechanical first law can be written as
\begin{eqnarray}
	\delta M=T_\text{H}\delta S_\text{BH}.\label{1st law}
\end{eqnarray}
Note that this equation is not obtained from only interpreting the black hole's quantities as the thermodynamic variables, but it is also purely derived from the geometric description. Another important thermodynamic variable is temperature. It can be obtained by considering the property of the black hole horizon, namely, the surface gravity $\kappa$, $T_\text{H}=\frac{\kappa}{2\pi}=\frac{1}{4\pi}\frac{\partial f(r)}{\partial r}\big|_{r=r_h}$. Also, the  Hawking temperature of the black hole can be obtained by using the thermodynamic description as follows:
\begin{eqnarray}
	T_\text{H}
	&=&\frac{\partial M}{\partial S_\text{BH}}
	=\frac{1+\Lambda r_h^2}{4\pi r_h}.
\end{eqnarray}
Now, let us apply the idea of the holographic principle to this kind of a black hole. According to Cohen's argument, the energy of the system in the quantum field theory never saturates to the energy of the black hole, $l^3 \Lambda_{UV}^4\lesssim l M^2_\text{P}$. As a result, one can obtain the energy density of dark energy by interpreting that it corresponds to one at the saturated point $\rho_{de} =\Lambda_{UV}^4 \sim  l^{-2}M^2_\text{P}$. The application of this energy density in cosmology is widely investigated in the literature, depending on identification of the cosmological length scale $l$. By following Cohen's argument, the energy density of dark energy is thus expressed as
\begin{eqnarray}
	\rho_{de}\delta V \sim \delta M_{BH}.
\end{eqnarray}
By requiring the validity of the first law in Eq.~\eqref{1st law} on the AdS black hole, the energy density of dark energy is then written as
\begin{eqnarray}
	\rho_{de}
	&\sim&T_\text{H}\frac{\delta S_\text{BH}}{\delta V}
	=\frac{1}{8\pi}\left(\frac{1}{L^2}+\Lambda\right).
\end{eqnarray}	
Note that the horizon radius of the black hole is now written as an arbitrary cosmological length scale, $L$. In addition, we have worked with the convention $G=1$ so that $M_\text{P}=\sqrt{1/8\pi}$. Conveniently, we can restore the Planck mass up to the dimensionless parameter $b$ as  
\begin{eqnarray}
	\rho_{de}=3M_\text{P}^2b^2\left(\frac{1}{L^2}+\Lambda\right). \label{rho-de}
\end{eqnarray}	
It is very important to note that the energy density in Eq.~\eqref{rho-de} can be also obtained from the dimensional estimation of the modification of Cohen's energy bound $L^3\Lambda_{UV}^4\lesssim M_\text{AdS}\sim M_\text{P}^2L(1+\Lambda L^2)$. The first term in the above expression corresponds to the energy density of the original HDE as investigated in Refs. \cite{Hsu:2004ri, Li:2004rb}.  Obviously, the result can reduce to the original HDE model by setting $\Lambda=0$. The additional term corresponds to a constant. Therefore, it provides a possibility of obtaining the late-time expansion of the Universe. Note that for positivity of the cosmological constant $\Lambda>0$, the black hole is asymptotically AdS. However, in cosmological context, the positive sign of $\Lambda$ corresponds to the asymptotic dS spacetime. This is the crucial key of the present work. There is a link between the AdS black hole (static and spherically symmetric solution) and the dS spacetime in cosmological context (homogenous and isotropic symmetry). In the next section, we will explicitly show how the Universe evolves described by this model of HDE.

\section{Cosmological models}\label{sec: Hubble hor} 
In the previous section, we obtain the general form of the energy density by using the idea of the holographic principle with the AdS black hole. In this section, we will specify the form of the energy density by adopting the length scale as the Hubble radius and the particle horizon. In order to investigate the possibility of obtaining the late-time expansion of the Universe, we consider the homogeneous and isotropic Universe with a flat Friedmann-Lema\^itre-Robertson-Walker (FLRW) metric,
\begin{eqnarray}
	\text{d}s^2=-\text{d}t^2+a(t)^2\left(\text{d}r^2+r^2\text{d}\theta^2+r^2\sin^2\theta\text{d}\phi^2\right),\label{FLRW-metric}
\end{eqnarray}
where $a(t)$ is a scale factor characterizing how the Universe evolves. We assume that the dynamics of the Universe is governed by the general relativity with the Einstein field equation 
\begin{eqnarray}
	G_{\mu\nu} = M_\text{P}^{-2} T_{\mu\nu},\label{EFE}
\end{eqnarray}
where $G_{\mu\nu}$ are components of the Einstein tensor and $T_{\mu\nu}$ are components of the energy-momentum tensor supposed to describe all contents in the Universe. In order to be compatible with the homogeneous and isotropic conditions, the form of the energy-momentum tensor can be considered as one for the perfect fluid which is the fluid with no heat transfer and viscosity. The energy-momentum tensor of the perfect fluid can be written in terms of the energy density $\rho$ and the pressure $p$ of the fluid as
\begin{eqnarray}
	T^\mu_{\hspace{.2cm}\nu}=\text{diag}\big(-\rho, p, p, p\big),\label{perfect-fluid}
\end{eqnarray}
where $\rho=\rho(t)$ and $p=p(t)$. Conveniently, let us consider the contents in the Universe composed of radiation, matter and dark energy. For these three contents, it is sufficient to characterize the possibility for obtaining the background evolution of the Universe. Using the metric in Eq.~\eqref{FLRW-metric} and the energy momentum for the three aforementioned contents, the $(0,0)$ and $(i,j)$ components of the Einstein field equation~\eqref{EFE} can be respectively expressed as
\begin{eqnarray}
	3M_\text{P}^2H^2
	&=&\rho_r+\rho_m+\rho_{de},\label{Ein 00}\\
	M_\text{P}^2\left(2\dot{H}+3H^2\right)&=&-p_r-p_m-p_{de}.\label{Ein ij}
\end{eqnarray}
where $H=\dot{a}/a$ is the Hubble parameter. The dot denotes the derivative with respect to the cosmic time $t$ and the quantities with subscripts ``$r$", ``$m$", and ``$de$" denote those for radiation, matter and dark energy, respectively. The energy density of dark energy can be obtained from the previous section while its pressure can be obtained by using the conservation equation $\nabla_\mu T^\mu_{\hspace{.2cm}\nu}=0$. The conservation equation for each content can be written in terms of its energy density and pressure as  
\begin{eqnarray}
	\dot{\rho}_i+3H(1+w_i)\rho_i&=&0.\label{cont eq}
\end{eqnarray}
Here, we have written the pressure for each content in terms of its equation of state parameter $w_i$ and the energy density as $p_i=w_i \rho_i$. Note that the equation of state parameters for the matter and radiation are fixed as $w_m =0$ and $w_r=1/3$, respectively. In order to determine the evolution of the Universe properly, let us define the dimensionless variable for each content as the density parameter by following
\begin{eqnarray}
	\Omega_i&=&\frac{\rho_i}{3M_\text{P}^2H^2}.
\end{eqnarray}
As a result, the field equation~\eqref{Ein 00} can be written in dimensionless form as
\begin{eqnarray}
	1=\Omega_r+\Omega_m+\Omega_{de}.
\end{eqnarray}
This equation is a constraint equation. Then, one can use it in order to eliminate one of the dynamical variables. The dynamical equations of density parameters can be written in terms of the autonomous system as follows:
\begin{eqnarray}
	\Omega_i' 
	= -3\Omega_i\left[(1+w_i)+\frac{2}{3}\frac{\dot{H}}{H^2}\right],
	\label{dOmega i}
\end{eqnarray} 
where the prime denotes the derivative with respect to $\ln a(t)$. Note that we have used the conservation equation in Eq.~\eqref{cont eq}. This equation can be thought of as the master equation in order to determine the dynamics of the background Universe. In terms of autonomous system, each equation is coupled with the function $2\dot{H}/3H^2$ which can be determined by using Eq.~\eqref{Ein 00} and Eq.~\eqref{Ein ij} as 
\begin{eqnarray}
	\frac{2}{3}\frac{\dot{H}}{H^2}
	=-\sum_i (1+w_i)\Omega_i.\label{Hd}
\end{eqnarray} 
By specifying the form of the energy density of the dark energy, $\rho_{de}$, one can determine the equation of state parameter $w_{de}$ as
\begin{eqnarray}
	w_{de}
	=-1-\frac{\dot{\rho}_{de}}{3H\rho_{de}}.\label{w de}
\end{eqnarray}
Once $w_{de}$ is determined, the other important quantity characterizing behavior of the Universe for each epoch is the effective equation of state parameter $w_{eff}$, 
\begin{eqnarray}
	w_{eff}
	=\frac{\sum_i p_i}{\sum_i \rho_i}
	=-1-\frac{2\dot{H}}{3H^2}.\label{w eff}
\end{eqnarray}
Now, we have the master equations and the main quantities to characterize the evolution of the Universe. This can be evaluated by specifying the length scale $L$ appearing in the energy density of dark energy in Eq.~\eqref{rho-de}. We will investigate the evolution of the Universe by choosing $L$ as Hubble radius and particle horizon in the next two subsections.

\subsection{Hubble radius as the length scale}
\begin{figure}[h!]
\begin{center}
	\includegraphics[scale=0.5]{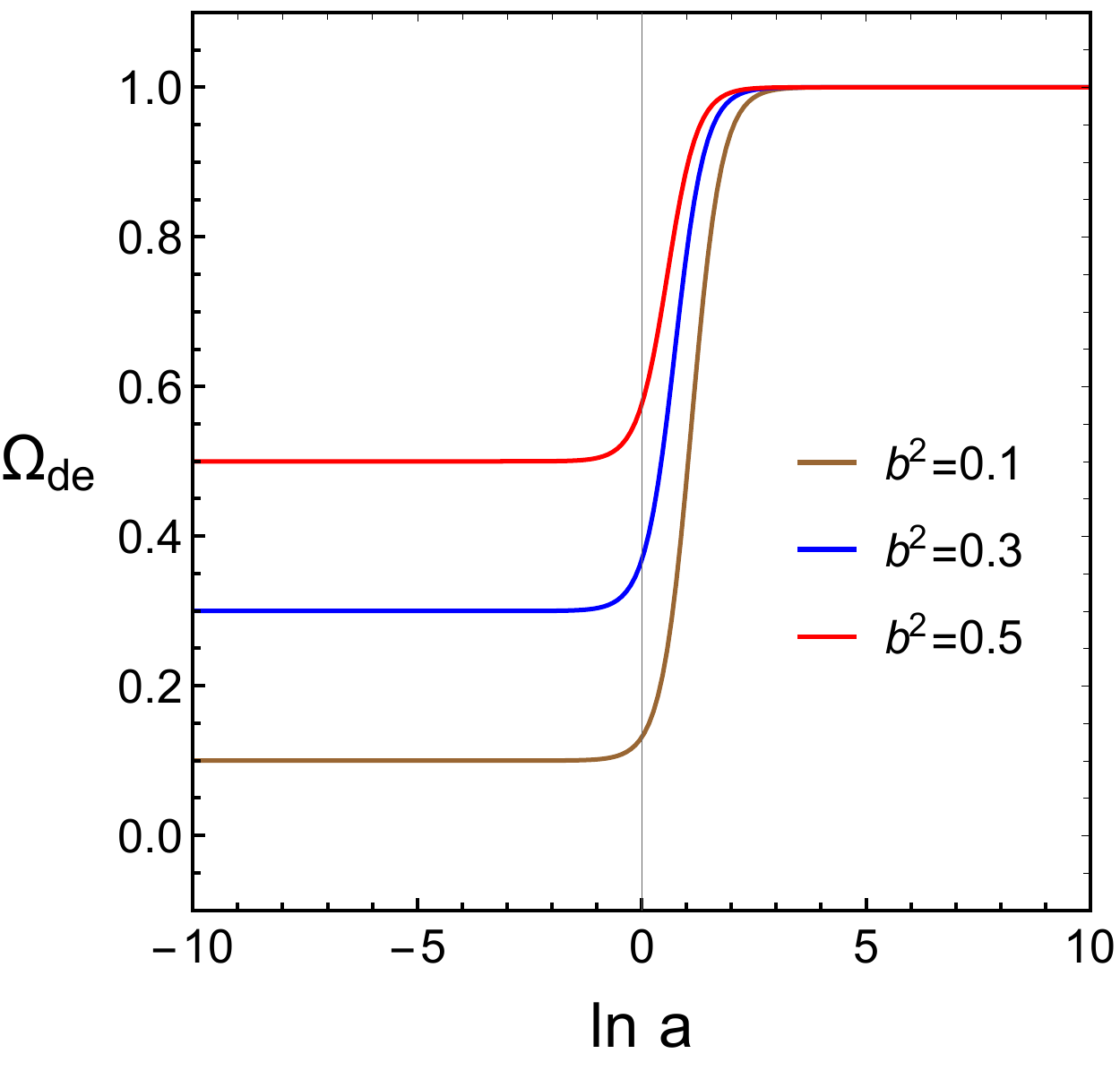}\hspace{1.cm}
	\includegraphics[scale=0.5]{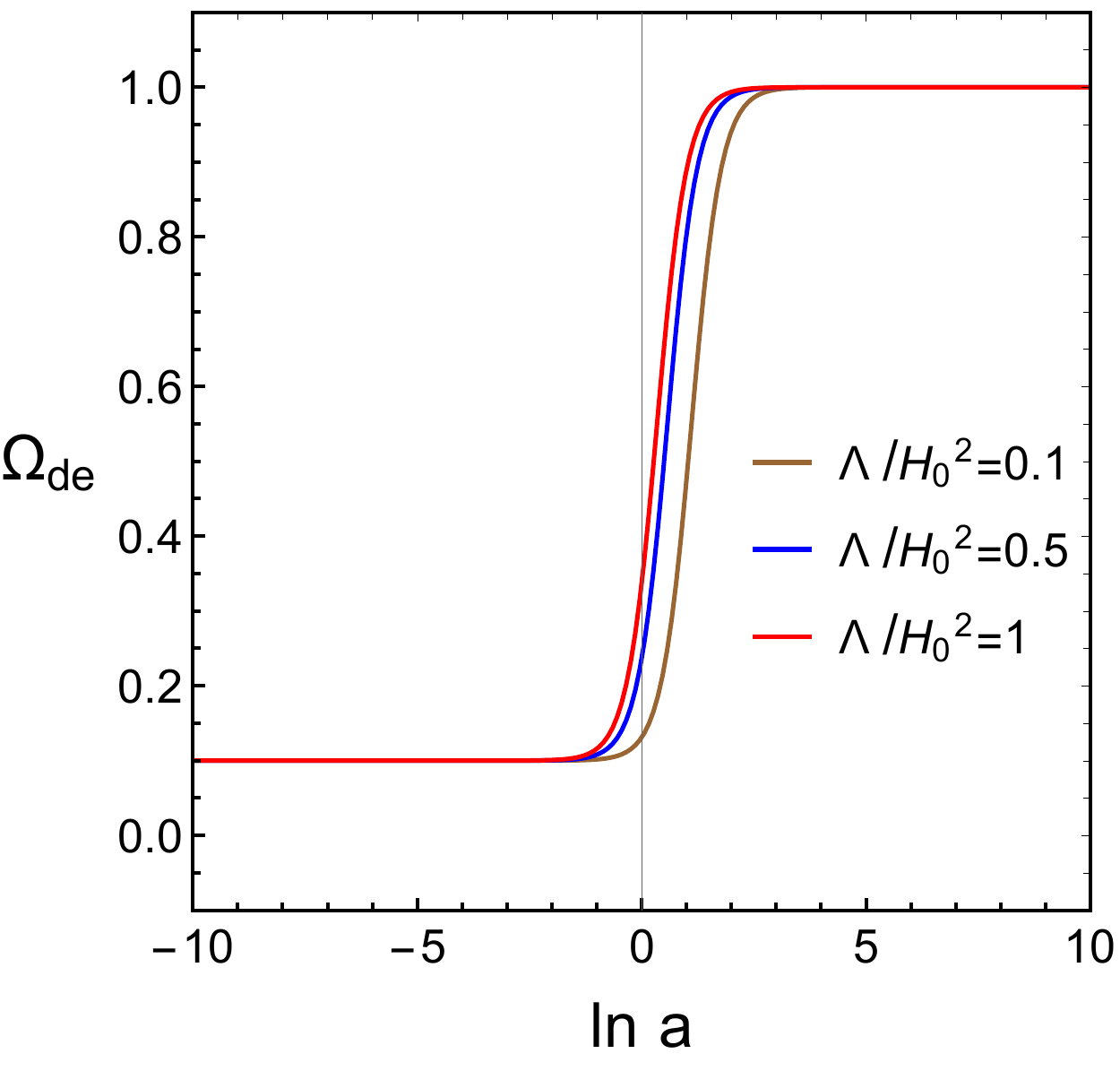}
\end{center}
\caption{The density parameter of the HDE with respect to $\ln a$ for various values of $b^2$ with fixing $\Lambda/H_0^2=0.1$ (left) and one for various values of $\Lambda/H_0^2$ with fixing $b^2=0.1$ (right). }\label{fig:Omega-Hubble}
\end{figure}

As we have seen from the beginning of this section, the dynamics of the Universe can be characterized by the evolution of the scale factor. Therefore, it is natural to find the cosmological length scale related to the scale factor. As a result, one of the simplest length scales can be characterized by the ratio of the scale factor and its time derivative, $a/\dot{a} = 1/H$ or the Hubble radius. Therefore, in this subsection, the length scale is chosen to be the Hubble radius given by
\begin{eqnarray}
	L=H^{-1}.
\end{eqnarray}
Substituting it to Eq.~\eqref{rho-de}, the energy density for HDE is then expressed as
\begin{eqnarray}
	\rho_{de}=3M_\text{P}^2b^2\Big(H^2+\Lambda\Big) = \rho_H + \rho_\Lambda.\label{rho Hubble}
\end{eqnarray}
From this expression, one can see that the energy density can be separated into two parts. The first part (i.e. $\rho_H$) is indeed a scaling solution. From Eq.~\eqref{w de}, the equation of state parameter for this part can be written as
\begin{eqnarray}
	w_{H}
	=-1-\frac{2\dot{H}}{3H^2} 
	= \sum_i w_i\Omega_i .\label{w H}
\end{eqnarray}
Obviously, it behaves like the dominant content in the Universe. Hence, $\rho_H$ corresponds to the original version of the HDE which is not possible to be responsible for the late-time expansion of the Universe \cite{Hsu:2004ri}. Since this part is the scaling solution, it may be dominating at an early epoch depending on the parameter $b$ as shown in the left panel in Fig.~\ref{fig:Omega-Hubble}. One of the strongest constraints on early dark energy comes from the big bang nucleosynthesis (BBN). This provides the upper bound on the density parameter of dark energy as $\Omega_{de} < 0.045$ \cite{Bean:2001wt}. As a result, in order to satisfy this condition, the parameter $b^2$ is constrained as
\begin{eqnarray}
	\Omega_H = b^2 < 0.045. \label{boundonb2}
\end{eqnarray}
For a conservative bound, we restrict our consideration as $b^2\sim 10^{-2}$.

For the second part, the energy density is a constant; therefore, it can play the role of dark energy driving the late-time expansion of the Universe. Even though it is the constant equivalent to the cosmological constant in the $\Lambda$CDM model, it is indeed rooted from the AdS black hole in the notion of the holographic principle. The behavior of the density parameter for various values of $\Lambda$ at the present epoch is shown in the right panel of Fig.~\ref{fig:Omega-Hubble}.

\begin{figure}[h!]
\begin{center}
	\includegraphics[scale=0.45]{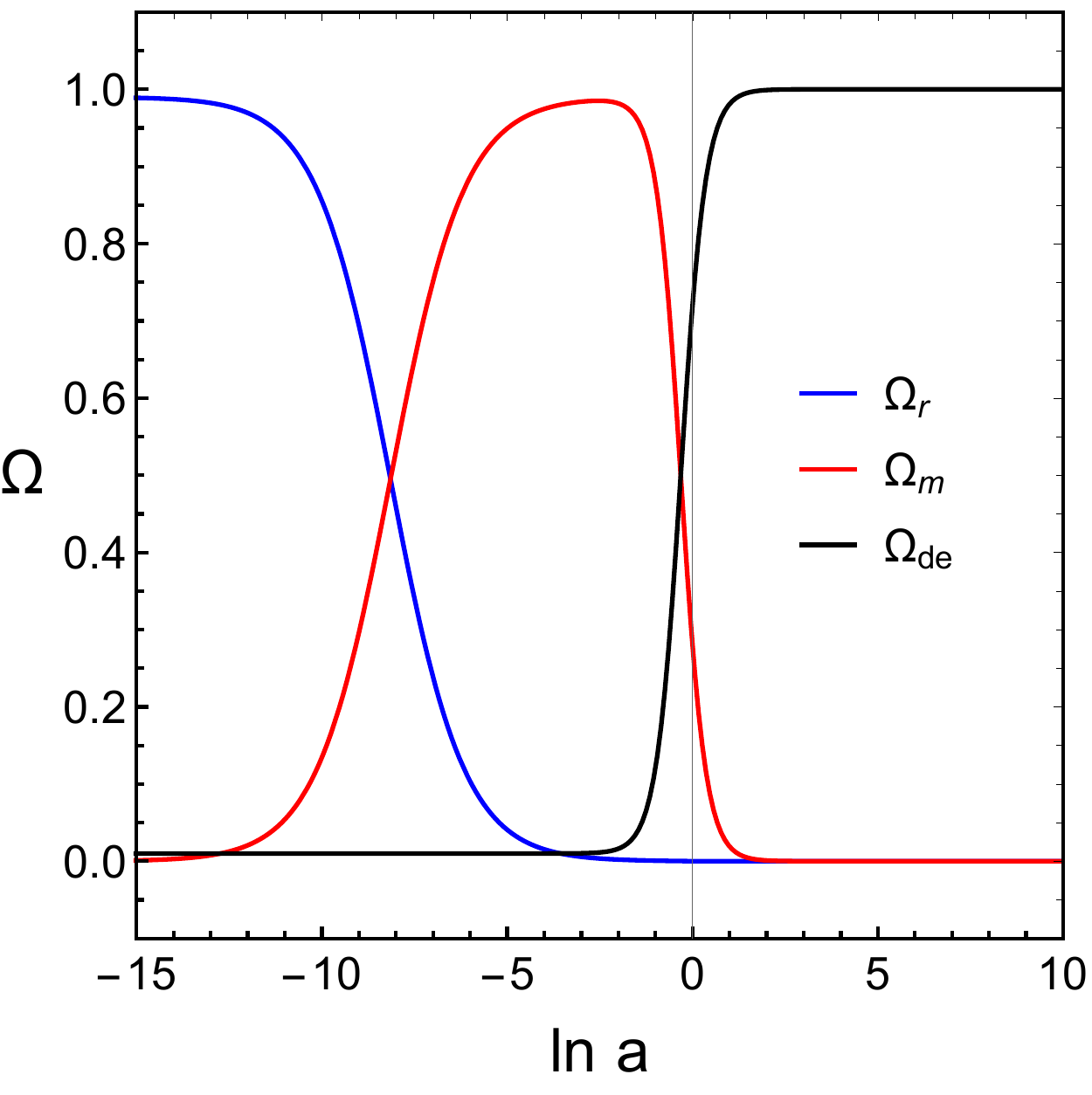}\hspace{1.cm}
	\includegraphics[scale=0.46]{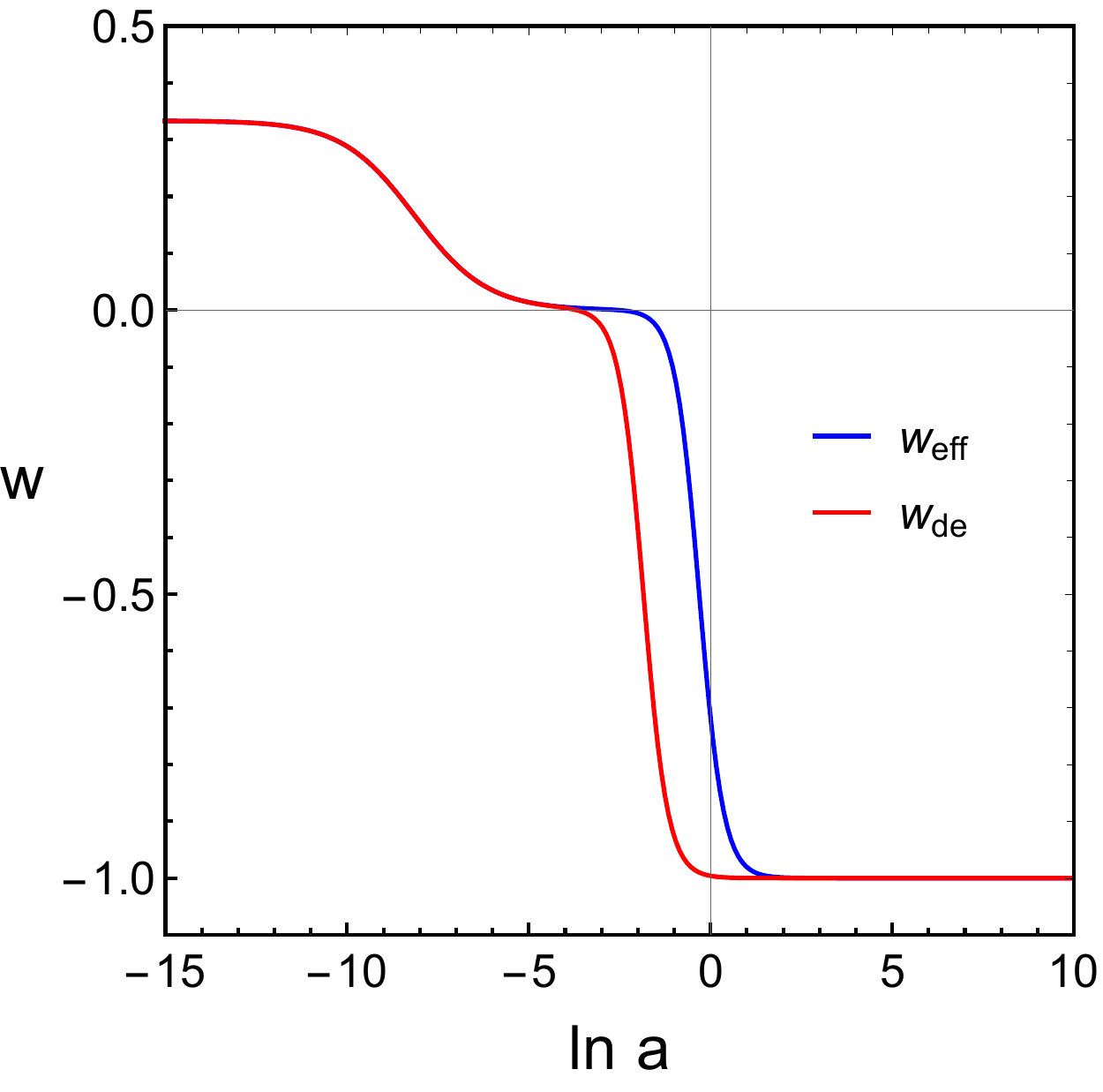}
\end{center}
\caption{The density parameters (left) and equation of state parameters (right) with respect to $\ln a$ for $b^2=10^{-2}$ and $\Lambda/H_0^2=71$.}\label{fig:Omega-Hubble2}
\end{figure}

One of the useful properties of this model is that one can solve exact solutions for all of its contents. This is due to the fact that the contribution from the proposed HDE is proportional to $H^2$ and a constant. In other words, one can algebraically solve equations for $H^2$ and $\dot{H}$ with a particular set of initial conditions without solving the differential equation. From Eqs.~\eqref{Ein 00} and \eqref{Ein ij}, one can rewrite them in terms of the current observed values as 
\begin{eqnarray}
	\frac{H^2}{H_0^2}
	&=&\frac{1}{1-b^2}\left(\Omega_{r,0}a^{-4}+\Omega_{m,0}a^{-3}+b^2\frac{\Lambda}{H_0^2}\right),\label{H^2 in a}
	\\
	\frac{2\dot{H}}{3H_0^2}
	&=&-\frac{1}{1-b^2}\left(\frac{4}{3}\Omega_{r,0}a^{-4}+\Omega_{m,0}a^{-3}\right).\label{Hdot in a}
\end{eqnarray}
Note that $\Omega_{r,0} = 8.1 \times 10^{-5}$, $\Omega_{m,0} = 0.28 -\Omega_{r,0}$, and $\Omega_{de,0} = 0.72$ are supposed to be the values of the present-time density parameters for radiation, matter, and dark energy, respectively. The value of $\Lambda/H_0^2$ is constrained by the current contribution of dark energy, i.e., $\Omega_{de,0} = 0.72$. As a result, we obtain $\Lambda/H_0^2=71$ for choosing $b^2=10^{-2}$. The Hubble parameter is known and the density parameter for each content can be written as 
\begin{eqnarray}
	\Omega_r&=&\frac{\Omega_{r,0}a^{-4}}{\frac{H^2}{H_0^2}},\hspace{1cm}
	\Omega_m=\frac{\Omega_{m,0}a^{-3}}{\frac{H^2}{H_0^2}},\hspace{1cm}
	\Omega_{de}	=b^2\left(1+\frac{\Lambda}{H^2}\right).	
\end{eqnarray}
As a result, the standard evolution can be obtained as shown in the left panel in Fig.~\ref{fig:Omega-Hubble2}. On the right panel, we also show that the effective equation of state parameter $w_{eff}$ evolves as the standard one. Moreover, one can see that $w_{de}$ is tracked on the dominant content which is a signature of the scaling solution contributed from the first part $\rho_H$ as we have discussed. Note also that $w_{eff}$ and $w_{de}$ can be written in terms of $\dot{H}$ and $H$ as follows:
\begin{eqnarray}
	w_{de}
	&=&-1-\frac{\frac{2\dot{H}}{3H_0^2}}{\frac{H^2}{H_0^2}+\frac{\Lambda}{H_0^2}},\\
	w_{eff}
	&=&-1-\frac{\frac{2\dot{H}}{3H_0^2}}{\frac{H^2}{H_0^2}}.
\end{eqnarray}

\subsection{Particle horizon as the length scale}\label{sec: partical hor} 
If the Universe has a finite age, the distance of light traveling from the past is limited. As a result, the distance at which we can receive the information from that point in the past is characterized by the particle horizon. In this subsection, we choose the cosmological length scale as the particle horizon defined as
\begin{eqnarray}
	R_h=a(t)\int_0^t\frac{\text{d}\bar{t}}{a(\bar{t}\,)}.
\end{eqnarray}
The energy density for this dark energy therefore reads
\begin{eqnarray}
	\rho_{de}&=&3M_\text{P}^2b^2\left(\frac{1}{R_h^2}+\Lambda\right) = \rho_R + \rho_\Lambda.\label{rho de with Rh}
\end{eqnarray}
As we have analyzed in the previous subsection, one can separate the contribution of the HDE into two parts, i.e., a constant contribution $\rho_\Lambda$ and a non-constant contribution $\rho_R$. For $\rho_\Lambda$, it is the constant and therefore plays the same role as that in the model with choosing the Hubble radius as the cosmological length scale. This part is responsible for driving the accelerated expansion of the Universe nowadays. Therefore, let us concentrate on $\rho_R = 3M^2_Pb^2/R^2_h$. In order to clarify how this part evolves, one can find the equation of state parameter by using Eq.~\eqref{w de} as follows:
\begin{eqnarray}
	w_{R}=-\frac{1}{3}\left(1-2 \frac{\Omega^{1/2}_R}{b}\right),\label{w R}
\end{eqnarray}
where $\Omega_R = \rho_R/(3M_P^2 H^2)$. From this expression, one can see that $w_R = -1/3$ during the radiation and matter dominated epochs. As a result, this part is not the scaling solution as found in the previous subsection. Therefore, the dimensionless parameter $b$ is not constrained in this case. We can set this parameter to be unity without loss of generality. From the fact that $w_R \sim -1/3$ during the matter dominated epoch, this provides a hint to us that it is possible to have fixed points such that $\Omega_R$ is dominated, since $\rho_R\sim a^{-3(1+w_R)} \sim a^{-2}$ decreases slower than $\rho_r$ and $\rho_m$.

In order to properly analyze the autonomous system, let us consider a general perfect fluid with the energy density $\rho_M$, the pressure $p_M$, and the equation of state parameter $w_M$. This fluid with subscript $``M"$ can be either radiation or matter as well as the total fluid. As a result, we have three dimensionless variables $\Omega_M$, $\Omega_R$, and $\Omega_\Lambda$ with one constraint,
\begin{eqnarray}
	1=\Omega_M+\Omega_\Lambda+\Omega_R.\label{Ein 00 Rh in Omega}
\end{eqnarray}
Therefore, we can use two dynamical equations as follows:
\begin{eqnarray}
	\Omega'_M
	&=&-3\Omega_M\left(1+w_M+\frac{2\dot{H}}{3H^2}\right),\label{Omega M prime}\\
	\Omega'_\Lambda
	&=&-3\Omega_\Lambda\left(\frac{2\dot{H}}{3H^2}\right).\label{Omega Lamb prime}
\end{eqnarray}  
From Eq.\eqref{Hd}, the term $2\dot{H}/(3H^2)$ can be  written as
\begin{eqnarray}
	\frac{2\dot{H}}{3H^2}
	&=&-(1+w_M)\Omega_M -(1+w_R) (1-\Omega_M).\label{Hdot}
\end{eqnarray}
According to the above result, the dynamical equations~\eqref{Omega M prime} and \eqref{Omega Lamb prime} are completely constructed as an autonomous system. The fixed points for this autonomous system can be computed from the following three conditions: (a) $\Omega_M=\Omega_\Lambda=0$, (b) $\Omega_M=\dot{H}=0$, and (c) $\Omega_\Lambda=1+w_M+\frac{2\dot{H}}{3H^2}=0$. Using Eqs.~\eqref{Ein 00 Rh in Omega} and \eqref{Hdot}, the fixed points are listed in Table~\ref{Tab:FP}. Here, we label the fixed points (a) and (b) for ones obtained from the conditions (a) and (b), respectively. For the fixed points obtained from the condition (c), they are labeled by (c1) and (c2).
\begin{table}
\centering
\caption{The density parameters and equation of state parameters for each fixed point.}
\begin{tabular}{| c | c | c | c | c | c |}
	\hline
\,\,Fixed point\,\,& $\Omega_M$ & \hspace{.5cm}$\Omega_\Lambda$\hspace{.5cm} & $\Omega_R$ & $w_{de}$ & $w_{eff}$ \\ 
 	\hline\hline
 (a) & 0 & 0 & 1 & $-\frac{1}{3}+\frac{2}{3|b|}$ & \,\,$-\frac{1}{3}+\frac{2}{3|b|}$\,\, \\
 	\hline  
 (b) & 0 & 1 & 0 & $-1$ & $-1$\\
 	\hline  
 (c1) & 1 & 0 & 0 &\,\,undefined\,\, & $w_M$\\
 	\hline  
 (c2) & $\,\,1-\frac{b^2}{4}(1+3w_M)^2$\,\, & 0 &\,\, $\frac{b^2}{4}(1+3w_M)^2$\,\, & $w_M$ & $w_M$\\
 	\hline    
\end{tabular}\label{Tab:FP}
\end{table}
Note that the effective equation of state parameter is found in Eq.~\eqref{w eff} with Eq.~\eqref{Hdot}. The equation of state parameter for dark energy can be written as
\begin{eqnarray}
	w_{de}
	&=&\frac{w_R \Omega_R + w_\Lambda \Omega_\Lambda}{\Omega_R + \Omega_\Lambda}.\label{w de 2}
\end{eqnarray} 
From this table, one can see that there exists fixed point (a) for which $\Omega_R$ is purely dominated as expected.

Using linear perturbation around the fixed points, one can identify their stabilities by considering the eigenvalues. If all eigenvalues are negative/positive, a fixed point is stable/unstable while if the eigenvalues are mixed between positive and negative, it is a saddle point. As a result, all stability conditions are listed in Table~\ref{Tab:stability}. From this table, one can see that fixed point (a) is unstable or saddle depending on the parameter $b$. Fixed point (b) corresponds to the $\Omega_\Lambda$-dominated epoch. Moreover, this fixed point is stable, then one can use this fixed point to be responsible for the late-time expansion of the Universe. Fixed point (c1) corresponds to radiation or matter dominated epoch. Naively, this fixed point is unstable as shown in Table~\ref{Tab:stability}. However, if we include the radiation and matter separately, it is found that this fixed point will be unstable and saddle for the radiation and matter dominated epochs, respectively. Then, there is a proper transition from the radiation dominated epoch to the matter dominated one. Note that if the fixed point for the matter domination is unstable, it is very difficult to tune the parameters in order to obtain the sufficient matter dominated epoch. Fixed point (c2) corresponds to the scaling solution, since the equation of state parameter for dark energy behaves similarly as that of the dominant content in the Universe, $w_{de} = w_{M}$. To obtain the positive density parameter, this fixed point exists when $|b| < 2/(1+3w_M)$. Then, this fixed point is a saddle point.
\begin{table}
\centering
\caption{The eigenvalues and stability for each fixed point.}
\begin{tabular}{| c | c | c |}
	\hline
\,\, Fixed point \,\,& Eigenvalues & Stability \\ 
 	\hline\hline
 (a) & $2+\frac{2}{\sqrt{b^2}},\,\,\frac{2}{\sqrt{b^2}}-(1+3w_M)$ & saddle when $|b|>2/(1+3w_M)$  \\
 	&	&\,\,unstable when $|b|<2/(1+3w_M)$\,\, \\
 	\hline  
 (b) & $-2,\,\,-3(1+w_M)$ & stable \\
 	\hline  
 (c1) & $3(1+w_M),\,\,1+3w_M$ & unstable \\
 	\hline  
 (c2) & \,\,$3(1+w_M),\,\,\frac{(1+3w_M)}{8}\Big[b^2(1+w_M)^2-4\Big]$\,\,& saddle because this point exists \\
 &	& only when $|b|<2/(1+3w_M)$ \\
 	\hline    
\end{tabular}\label{Tab:stability}
\end{table}

\begin{figure}[h!]
\begin{center}
	\includegraphics[scale=0.5]{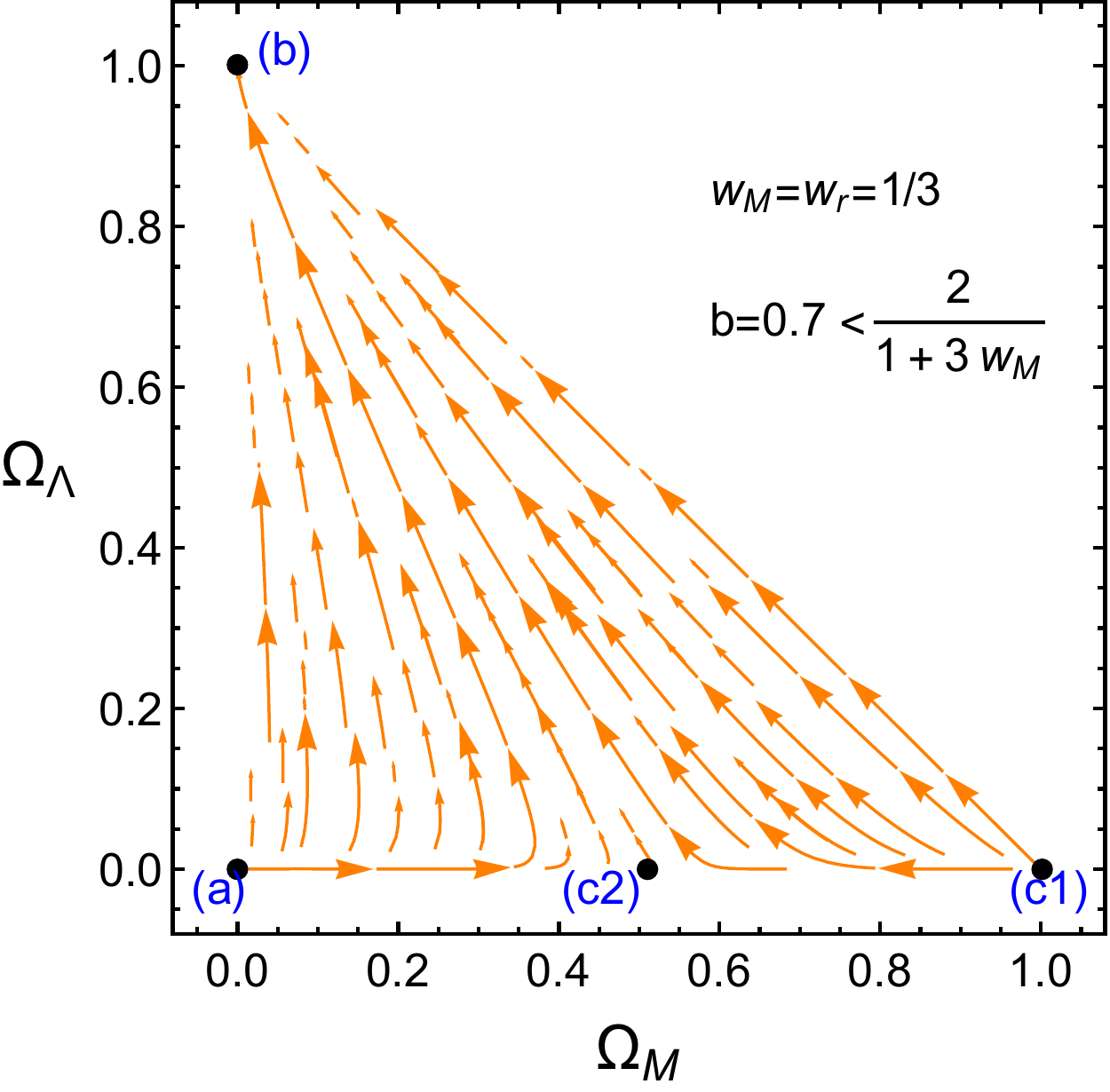}\hspace{1.cm}
	\includegraphics[scale=0.5]{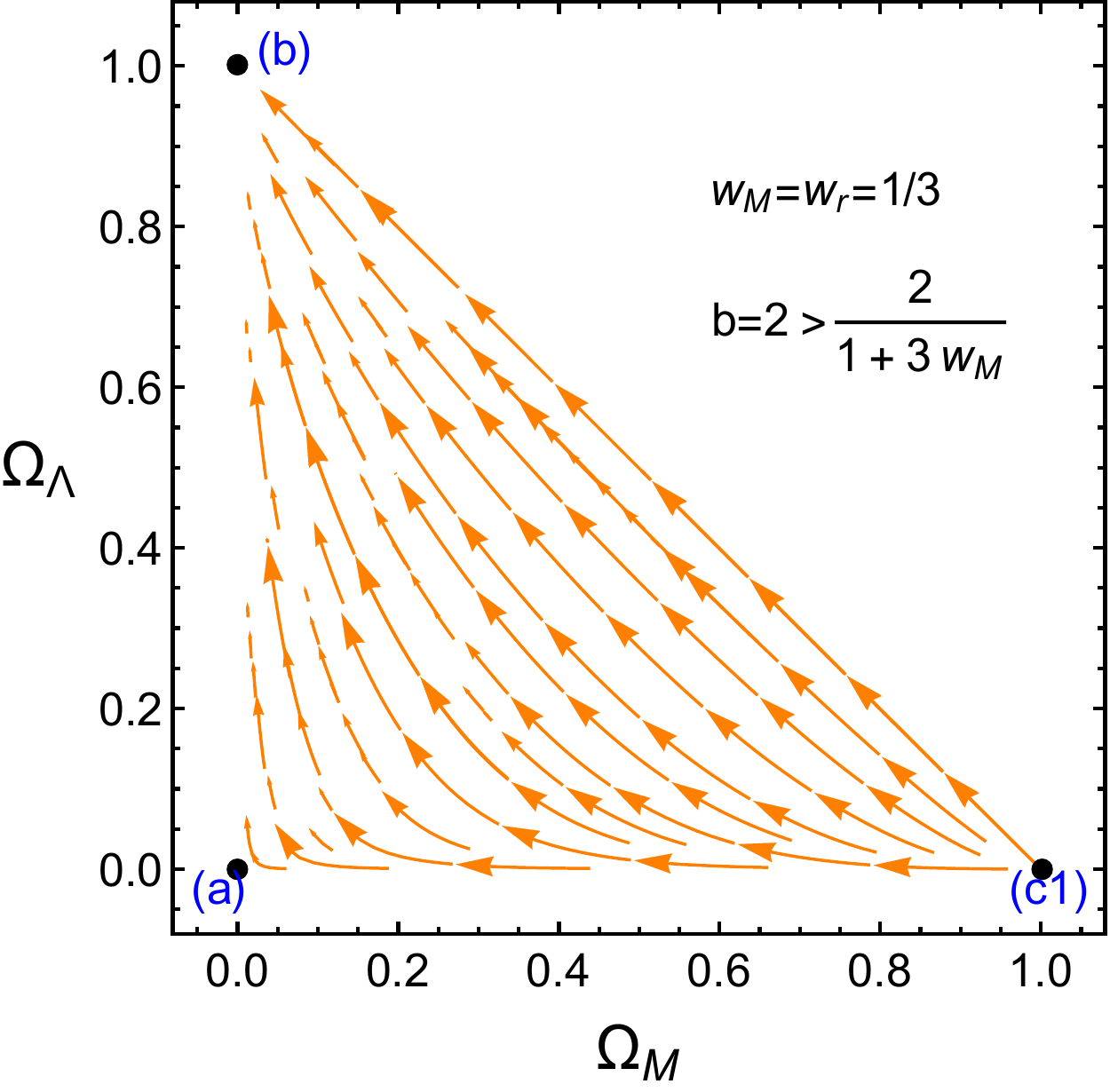}\\\vspace{.5cm}
	\includegraphics[scale=0.5]{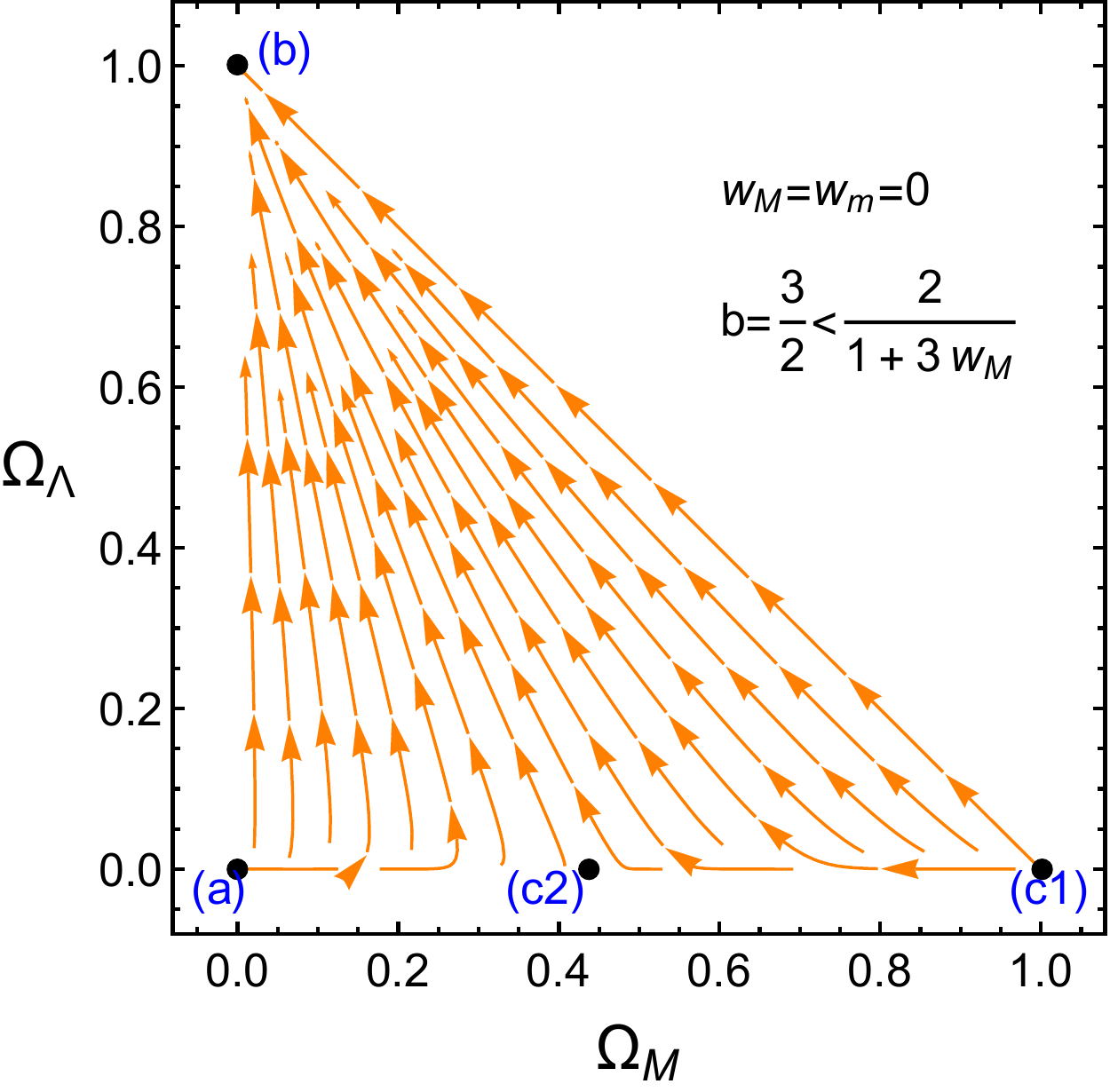}\hspace{1.cm}
	\includegraphics[scale=0.5]{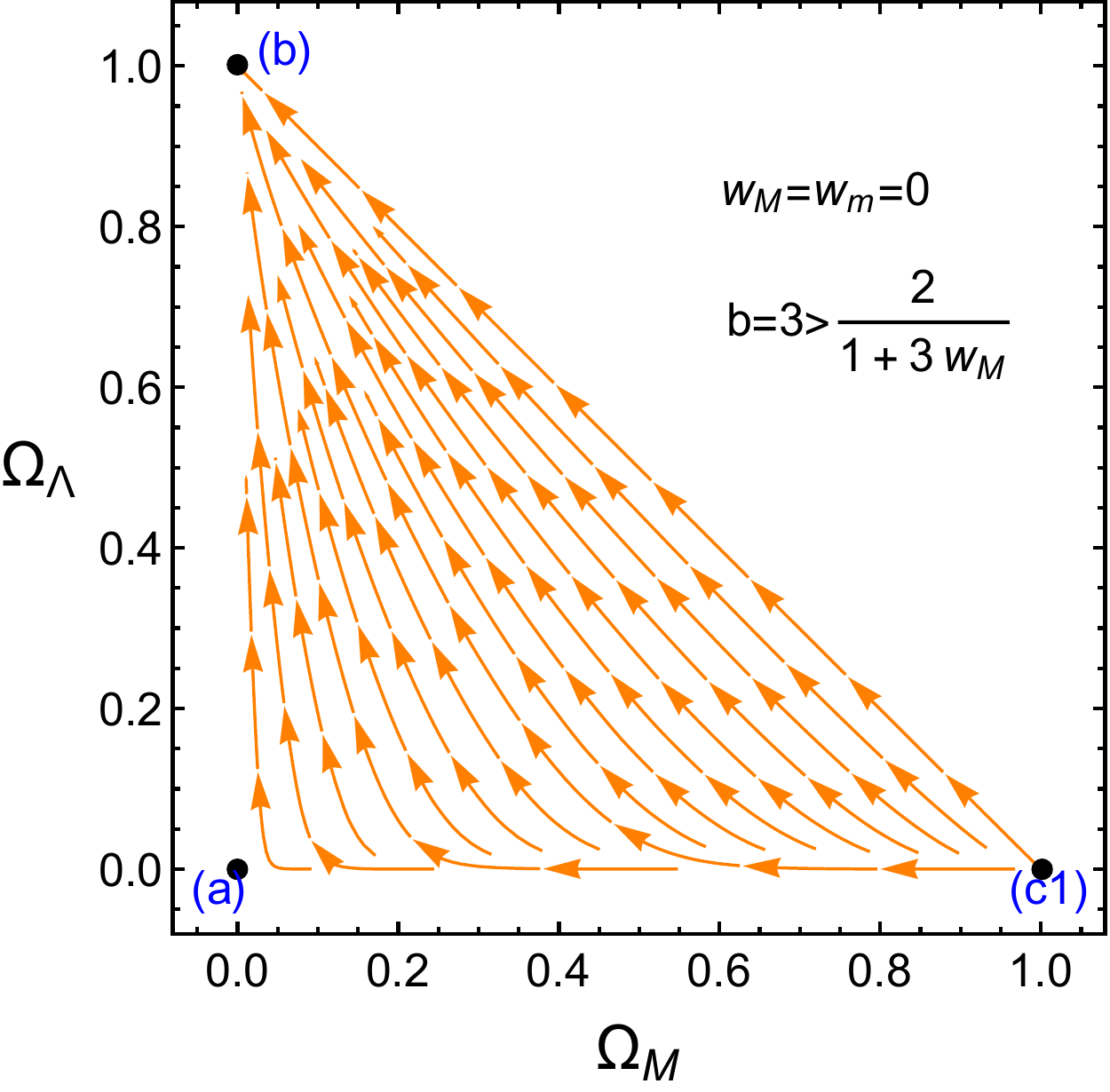}
\end{center}
\vspace{-.5cm}
\caption{The evolutions of the system passing through the fixed points when $|b| < 2/(1+3w_M)$ (left column) and $|b| > 2/(1+3w_M)$ (right column).}\label{evo1}
\end{figure}

From this analysis, one can separate the evolution of the Universe into two classes; $|b| < 2/(1+3w_M)$ and $|b| > 2/(1+3w_M)$. Note that, for $|b| = 2/(1+3w_M)$, fixed point (c2) is identical to fixed point (a). For the case $|b| < 2/(1+3w_M)$, fixed points (a) and (c2) are unstable and saddle, respectively. Therefore, it is possible to transit from fixed point (c1) corresponding to the radiation/matter dominated epoch to fixed point (c2) and then evolve further to the stable fixed point (b) corresponding to the $\Omega_\Lambda$-dominated epoch as shown in the left column in Fig.~\ref{evo1}. For the right column in this figure, the evolutions correspond to the case $|b| > 2/(1+3w_M)$. In this case, fixed point (a) is saddle while fixed point (c2) does not exist. Therefore, the system evolves from fixed point (c1) corresponding to  radiation/matter dominated epoch to fixed point (a) and then evolves further to the stable fixed point (b). In this figure, we separately consider the contribution for the radiation (above row) and matter (below row). Note that the system can evolve directly from fixed point (c1) to fixed point (b) depending on how much the density parameter $\Omega_{R,0}$ is nowadays. If we set $\Omega_{R,0}=0$, the system goes along the straight path from fixed point (c1) to fixed point (b). This setting is associated with $\Omega_R$ being absent for all time of evolution. Hence, this is indeed the cosmic evolution for the $\Lambda$CDM model. Furthermore, for $\Omega_{R,0}\neq0$, there exists a bubble contribution of $\Omega_R$ during the matter dominated and dark energy dominated epochs. 

\begin{figure}[h!]
\begin{center}
	\includegraphics[scale=0.53]{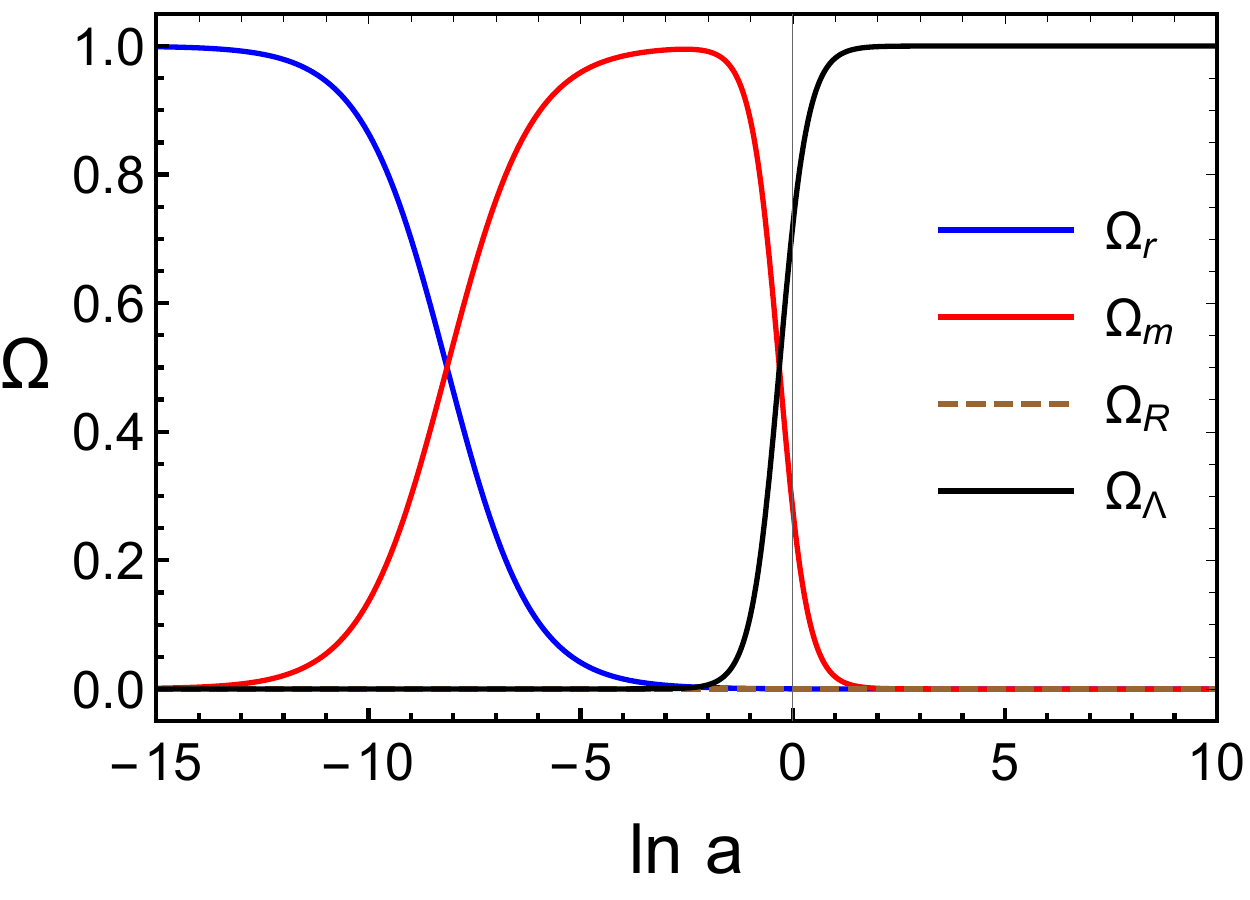}\hspace{1.cm}
	\includegraphics[scale=0.54]{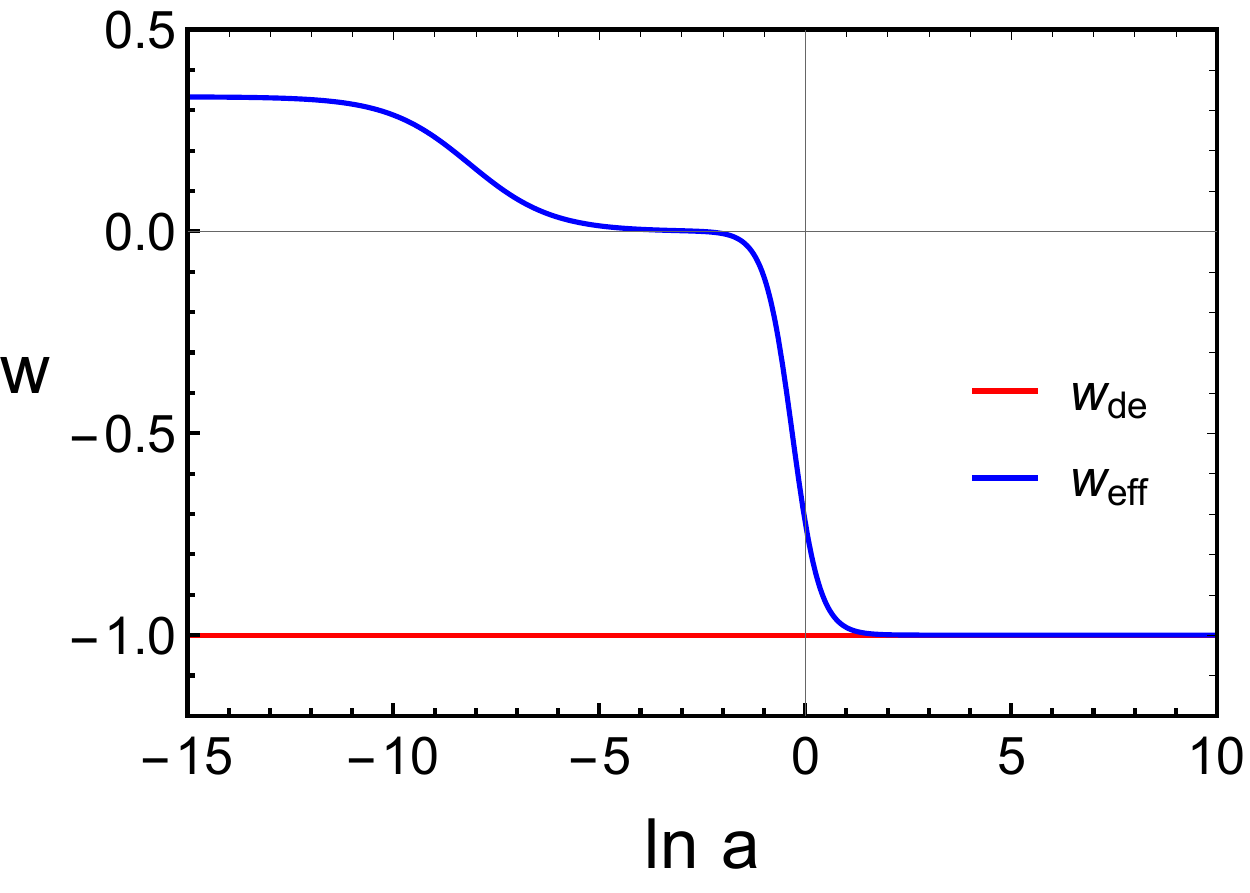}\\
\vspace{.5cm}
	\includegraphics[scale=0.53]{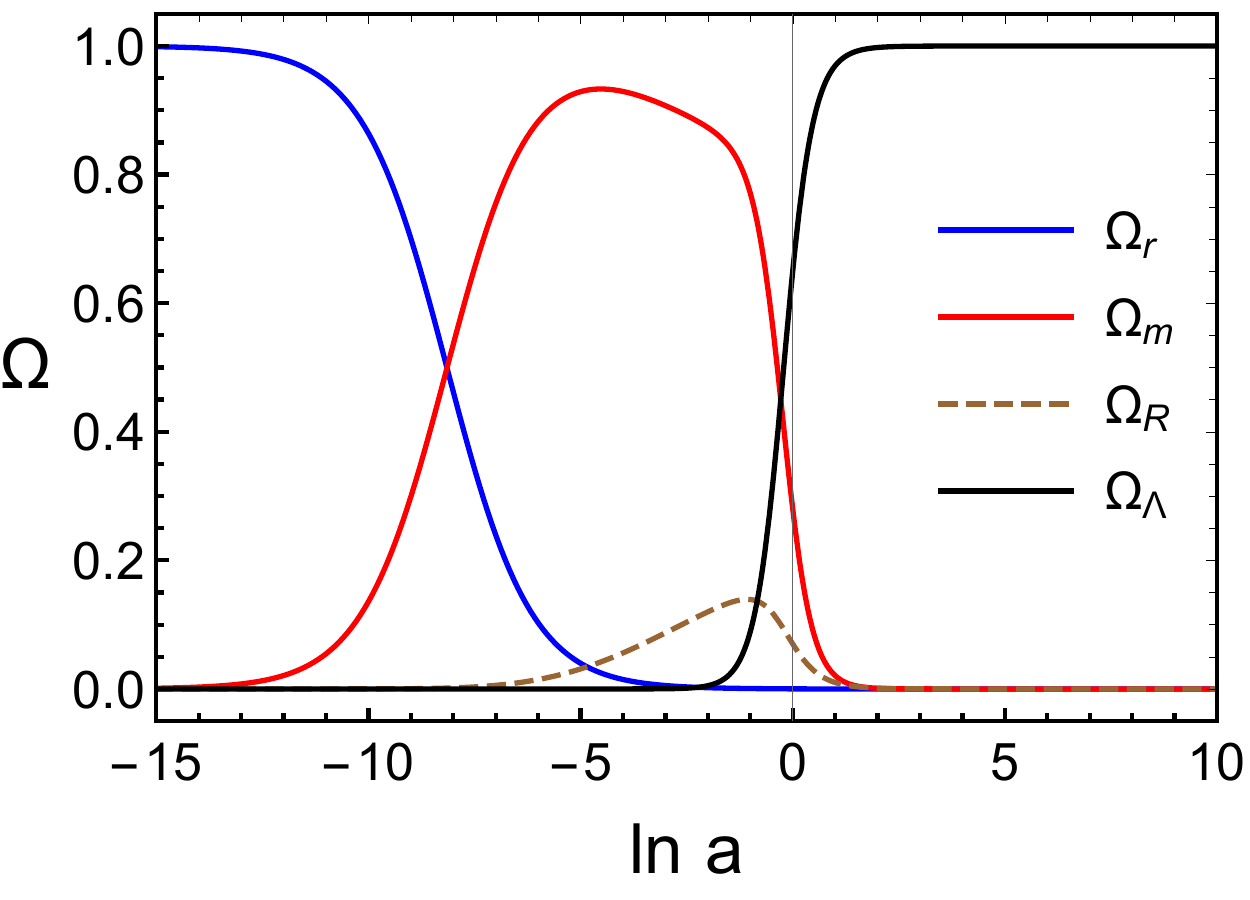}\hspace{1.cm}
	\includegraphics[scale=0.54]{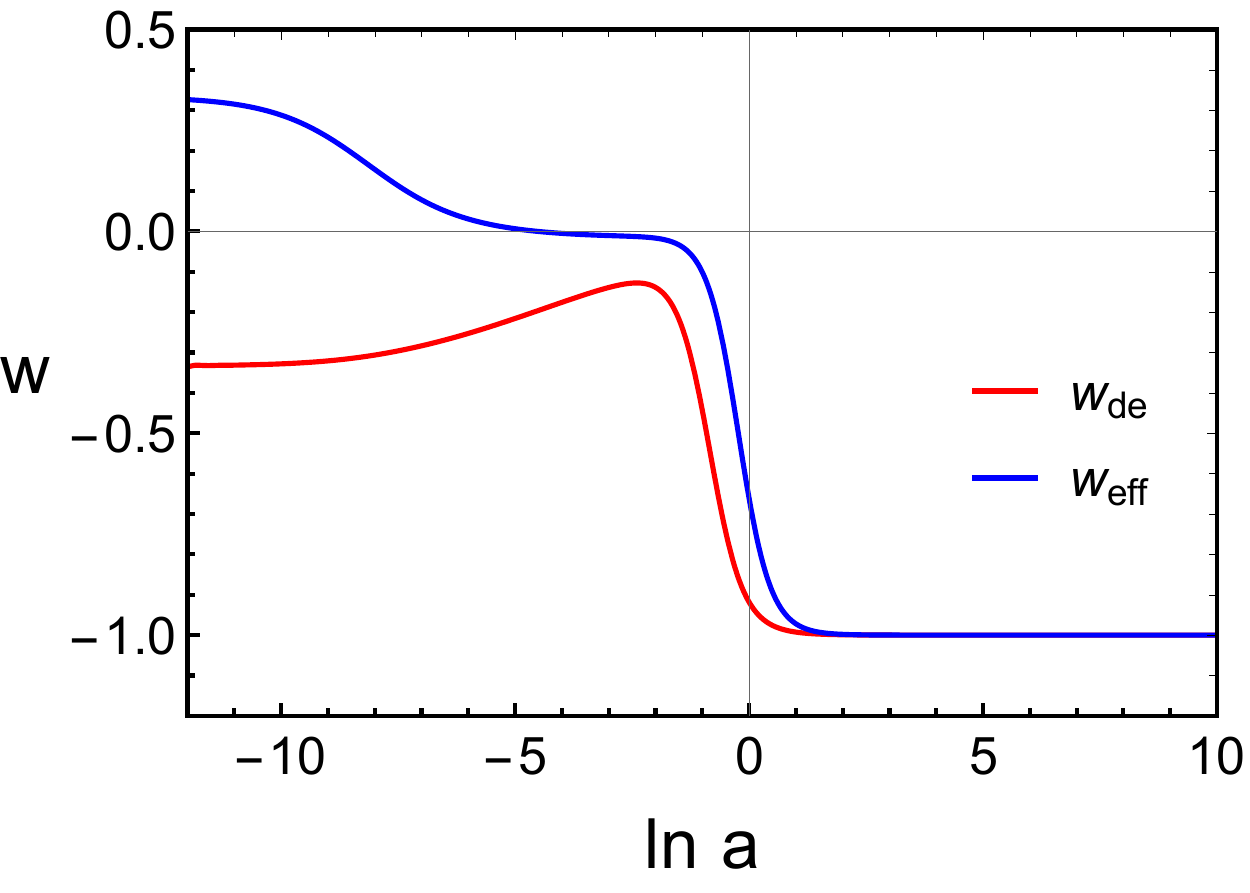}\\
\vspace{.5cm}
	\includegraphics[scale=0.53]{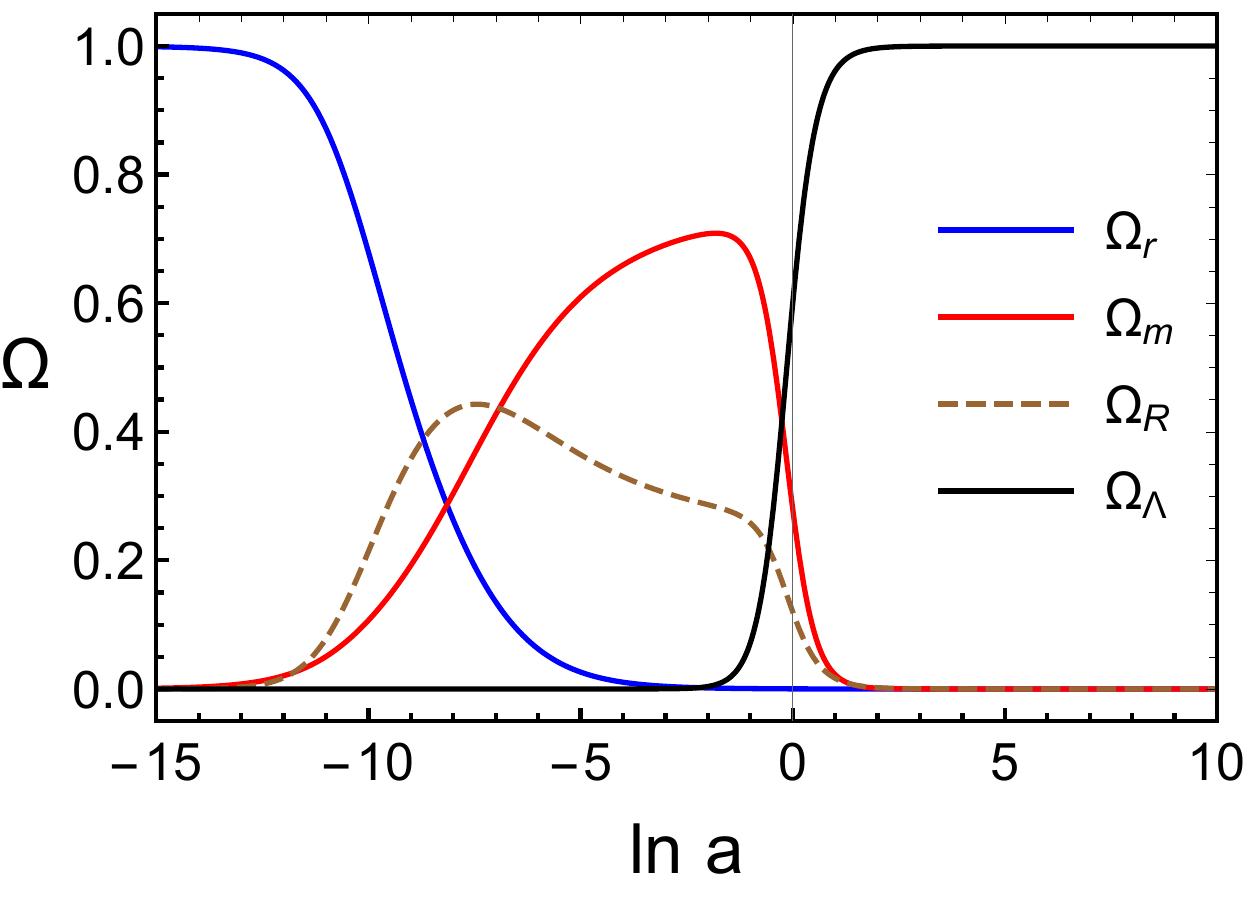}\hspace{1.cm}
	\includegraphics[scale=0.54]{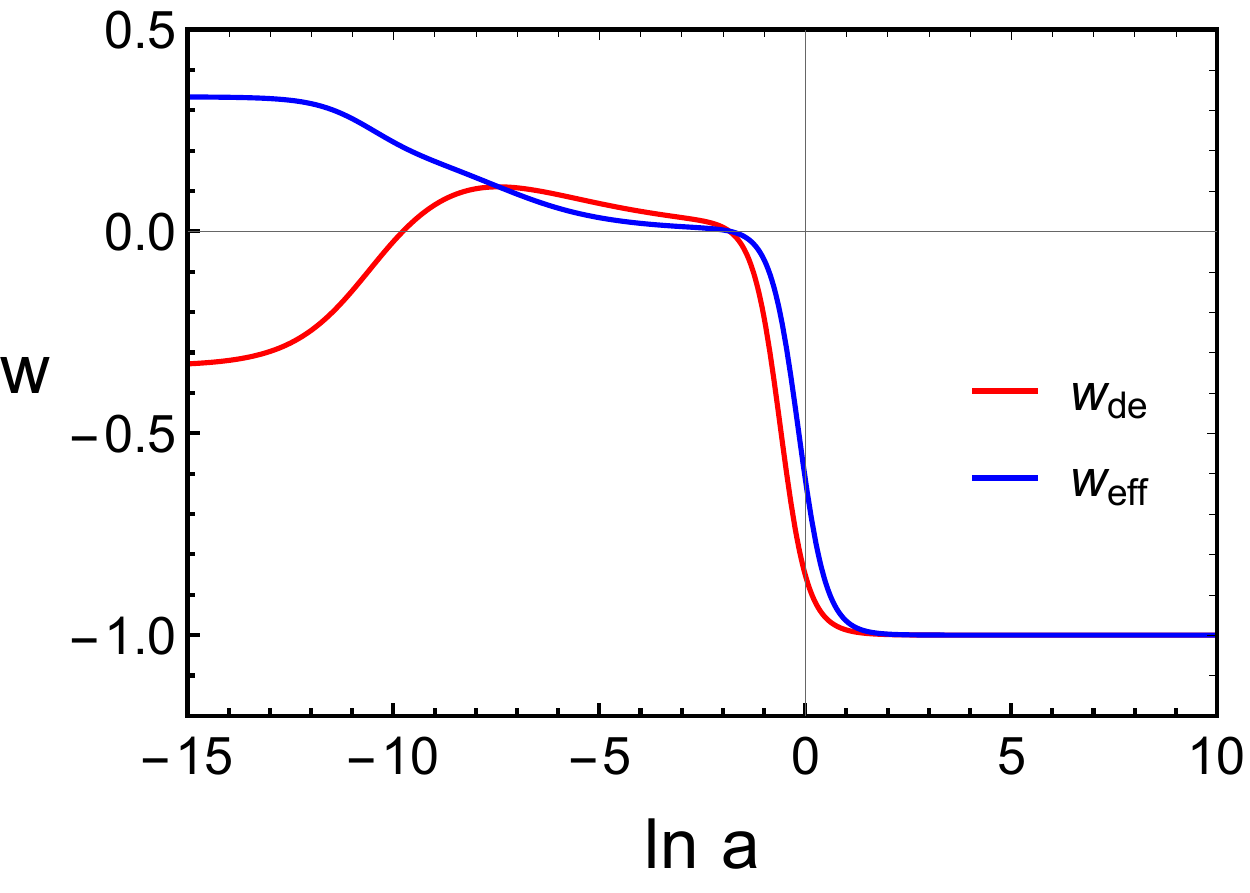}
\end{center}
\caption{Left column: evolutions of density parameters for radiation, matter and two parts for dark energy for various values of $\Omega_{R,0}$. Right column: evolutions of equation of state parameters for HDE and those of effective ones. Here, $\Omega_{R,0}=0$ (first row), $\Omega_{R,0}=0.07$ (second row), and $\Omega_{R,0}=0.12$ (third row) with fixing $b=1$.}\label{evo2}
\end{figure}

It is important to note that, from Eq.~\eqref{w de 2}, $w_{de}$ during the radiation and matter dominated epochs seems to be divergent, since $\Omega_{de}=\Omega_R+\Omega_\Lambda=0$. However, one can rearrange the expression in Eq.~\eqref{w de 2} to obtain the finite value for $w_{de}$ during the radiation and matter dominated epochs as follows:
\begin{eqnarray}
	w_{de}
	&=&\frac{w_R}{1+\Omega_\Lambda/\Omega_R}+\frac{w_\Lambda} {\Omega_R/\Omega_\Lambda+1}\sim w_R =-\frac{1}{3},
\end{eqnarray} 
where we have used the fact that during the radiation and matter dominated epochs, the ratio $\Omega_\Lambda/\Omega_R \propto a^2\sim 0$. This is also compatible with the numerical results as shown in the right panels of the second and third rows in Fig.~\ref{evo2}. Note also that we choose the initial conditions describing the density parameters at the present time as $\Omega_{r,0}=8.1\times10^{-5}, \Omega_{m,0}=0.28-\Omega_{r,0}$ and $\Omega_{\Lambda,0}=0.72-\Omega_{R,0}$.

Let us discuss the equivalence between the dynamical system and numerical approaches in details. For the first row in Fig.~\ref{evo2} corresponding to the case $\Omega_R$ being absent, one can see that the evolution of the Universe is actually the same as the one predicted by the standard $\Lambda$CDM model. This corresponds to the system evolving along the path of the fixed points (c1) $\to$ (b) as mentioned previously.  By increasing $\Omega_{R,0}$, the amount of the matter in its own dominated epoch will decrease. Hence, from the numerical results, we can conclude that if the amount of $\Omega_{R,0}$ does not vanish (this can be illustrated in the second and third rows in Fig.~\ref{evo2}), the evolution of the Universe will follow the path of the fixed points (c1) $\to$ (c2) $\to$ (b). As seen in the left column in Fig.~\ref{evo1}, there exists the evolution path (a) $\to$ (c2) $\to$ (b). Note also that by varying the parameter $b$, it does not affect the overall feature of the cosmic evolution. Since the path (a) $\to$ (c2) $\to$ (b) corresponds to the cosmic evolution without the radiation dominated epoch, we do not consider this case to be representative of the observable Universe.

\section{Observational constraints}\label{sec: Constraint} 
We perform the likelihood analysis on the model parameters using the publicly available Markov chain Monte Carlo (MCMC) code \texttt{SimpleMC}\footnote{\href{https://igomezv.github.io/SimpleMC/index.html}{https://igomezv.github.io/SimpleMC/index.html}} The observation datasets include:
\begin{itemize}
	\item Type Ia supernova, with the binned Pantheon dataset \cite{Pan-STARRS1:2017jku};
	\item Cosmic chronometers approach on BOSS datasets \cite{Moresco:2016mzx}, where the galaxy ages are used to constrain the expansion history of the Universe \cite{Jimenez:2001gg,Schlegel:2009hj,SDSS:2011jap,BOSS:2012dmf};
	\item Baryon acoustic oscillations: CBAO (new version of BAO) including DR12Consensus \cite{BOSS:2016wmc,deSainteAgathe:2019voe,Blomqvist:2019rah,2011MNRAS.416.3017B,Ross:2014qpa,Ata:2017dya}.
\end{itemize}
The Metropolis--Hastings algorithm has been used to draw samples for MCMC analysis. We also use Akaike information criterion (AIC) to determine the quality of fits where the increasing number of parameters are penalized. With the $k$ parameters and the highest likelihood $\mathcal{L}_{\rm max}$, AIC is defined as
\begin{equation}
    {\rm AIC} = -2\ln \mathcal{L}_{\rm max} + 2k.
\end{equation}
Note that the last term is introduced in order to prevent an over-fitting issue. Comparing the two models, the difference between the estimators can be used to justify the statistical compatibility between them. Using Jeffreys' scale  \cite{1939thpr.bookJ}, the value of $\Delta$AIC $\leq 2$ implies statistical compatible, where the condition $2 < \Delta$AIC $\leq 5$ means there is mild tension between the models. The high statistically significant tension between models happens when $\Delta$AIC $> 5$. Gelman--Rubin diagnostic (GRstop) has been used for checking the convergence. In this section, we consider two scenarios in which the Hubble radius is used as the length scale (AdS-HDE: $L = H^{-1}$) and the particle horizon is used as the length scale (AdS-HDE: $L = R_h$), respectively. For each scenario, the parameter space of
\begin{align}
    &\text{AdS-HDE:} \;\; L = H^{-1}:\quad \{\Omega_{M,0}, \Omega_{b,0}, h, b^2\}\\
    &\text{AdS-HDE:} \;\; L = R_h:\quad \{\Omega_{R,0}, \Omega_{M,0}, \Omega_{b,0}, h, b^2\}
\end{align}
have been sampled, where $\Omega_{R,0}$ is the density parameter of dark energy contributed from the non-constant part at the present time, $\Omega_{M,0}$ is the total matter density parameter for other species combined (radiation, matter, and neutrino) excluding dark energy at the present time, $\Omega_{b,0}$ is the present-time density parameter for the baryon, and $h=H_0/100$ is the scaled Hubble scale.

\subsection{Hubble radius as the length scale ($L=H^{-1}$)}
\begin{figure}[h!]
\begin{center}
	\includegraphics[scale=0.43]{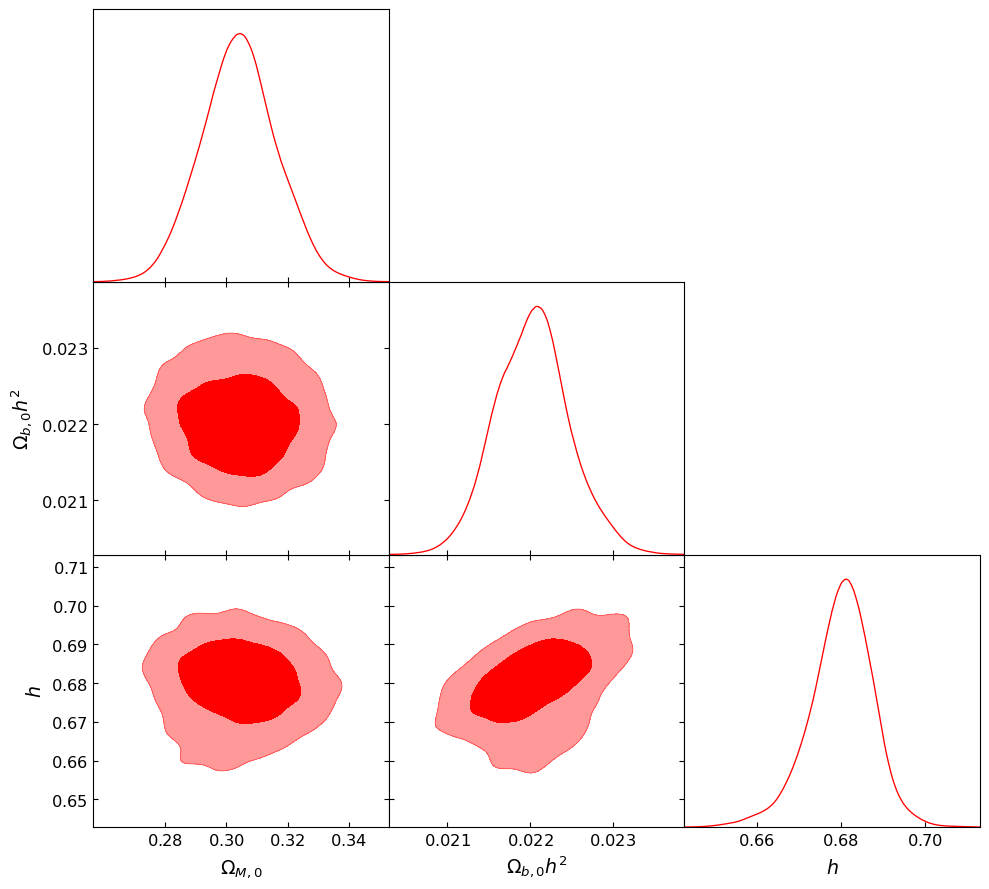}
	\includegraphics[scale=0.51]{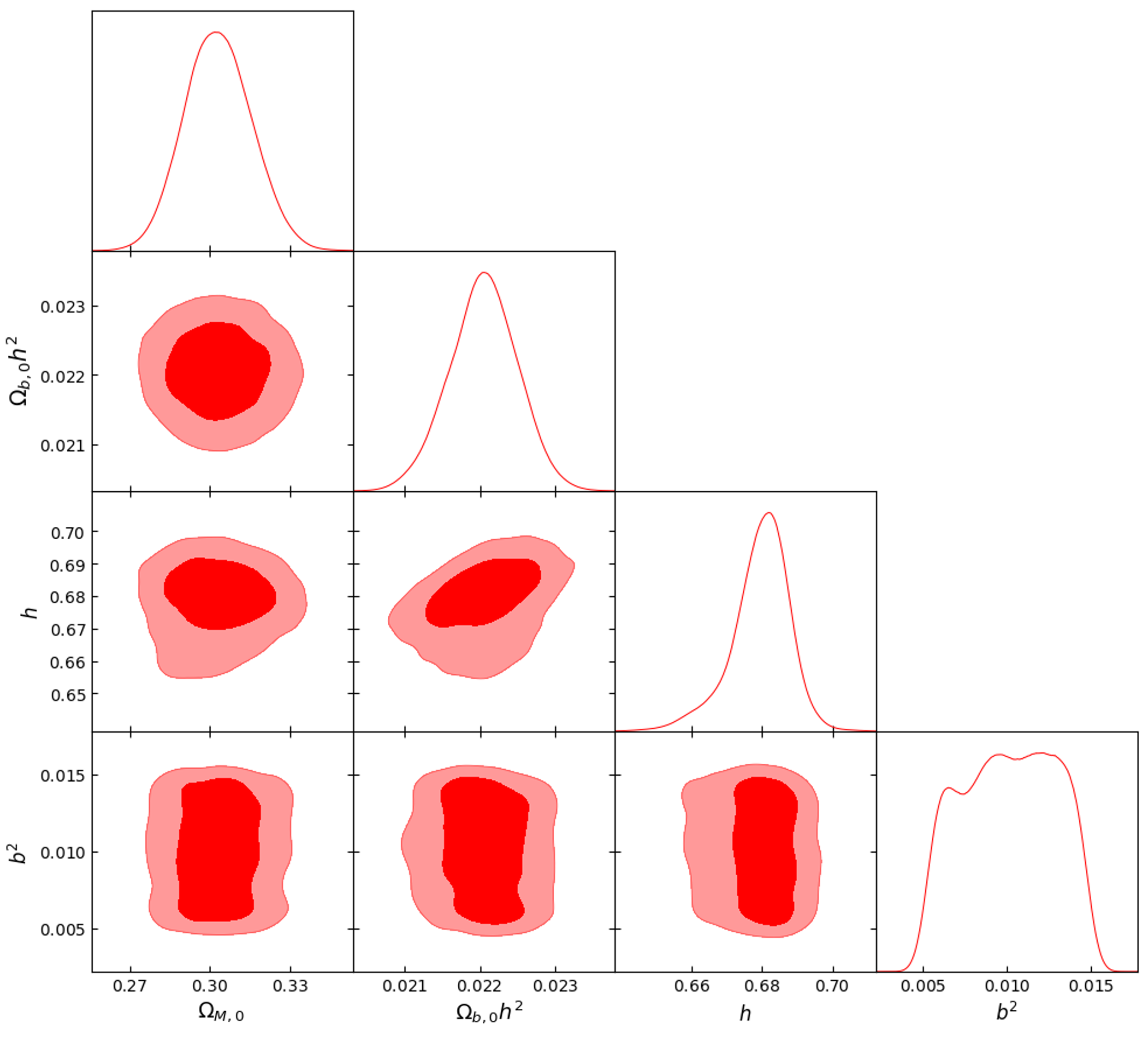}
\end{center}
\caption{Two-dimensional 68\% and 95\% C.L. allowed regions and the probability posterior distributions for model parameters from the AdS-HDE: $L = H^{-1}$ model. In the top panel, the value of $b^2=0.01$ is chosen to be consistent with the constraint~\eqref{boundonb2}. The analysis for the full parameter space $\{\Omega_{M,0}, \Omega_{b,0}, h, b^2\}$ is in the bottom panel.}\label{H-contour}
\end{figure}

Since each of the conventionally defined density parameters can be thought of as being scaled by $1\big/(1-b^2)$ as shown in Eq.~\eqref{H^2 in a}, the observation values for the total density parameter for each species in this model are interpreted with the relation
\begin{equation}
    \Omega_{i,0}^{\rm obs} = \frac{\Omega_{i,0}}{1-b^2}.
\end{equation}
The observational constraint analysis is reported in Fig.~\ref{H-contour}. The plot on the top panel shows the marginalized contours for the case that $b^2$ is fixed to be 0.01 for the conservative bound considered in Eq.~\eqref{boundonb2}. The marginalized contours for the full analysis of the model is on the bottom panel. The central values of the allowed region of the model parameters are given in Table~\ref{Tab: model parameters-Hubble}. For such case, we obtain the observed values:
\begin{equation}
    \Omega_{M,0}^{\rm obs} = 0.3062^{+0.0184}_{-0.0175},\quad \Omega_{b,0}^{\rm obs} = 0.0222^{+0.0006}_{-0.0006},\quad h = 0.6791^{+0.0110}_{-0.0127}.
\end{equation}
For the full analysis of $L = H^{-1}$ model (including fitting of $b^2$), we obtain the observed values:
\begin{equation}
    \Omega_{M,0}^{\rm obs} = 0.3056^{+0.1184}_{-0.1144},\quad \Omega_{b,0}^{\rm obs} = 0.0223^{+0.0079}_{-0.0077},\quad 
    h = 0.6700^{+0.0118}_{-0.0145},\quad 
    b^2 = 0.0098^{+0.0032}_{-0.0031}.
\end{equation}


\begin{table}
\centering
\caption{Observational constraints on the parameters of AdS-HDE: $L = H^{-1}$ using binned Pantheon, cosmic chronometers and baryon acoustic oscillations datasets}
\begin{tabular}{|l|c|c|c|c|}
	\hline
Models & $\Omega_{M,0}$ & $\Omega_{b,0}h^2$ & $h$ & $b^2$ 
    \\\hline\hline
AdS-HDE: $L=H^{-1}$, $b^2=0.01$ & $0.3031^{+0.0182}_{-0.0173}$ & $0.0220^{+0.0006}_{-0.0006}$ & $0.6791^{+0.0110}_{-0.0127}$ & 0.01 
    \\\hline  
AdS-HDE: $L=H^{-1}$ & $0.3026^{+0.0185}_{-0.0176}$ & $0.0221^{+0.0006}_{-0.0006}$ & $0.6778^{+0.0118}_{-0.0145}$ & $0.0098^{+0.0032}_{-0.0031}$
    \\\hline
$\Lambda$CDM & $0.3049^{+0.0184}_{-0.0179}$ & $0.0220^{+0.0006}_{-0.0006}$ & $0.6819^{+0.0117}_{-0.0138}$ & -- 
    \\\hline	
\end{tabular}\label{Tab: model parameters-Hubble}
\end{table}

Comparing with the $\Lambda$CDM in Table~\ref{Tab: model parameters-Hubble}, one can clearly see that the central values for both analyses stay inside the 1$\sigma$ regions of the $\Lambda$CDM. Although choosing $b^2 = 0.01$ gives a better estimator than that of $\Lambda$CDM, the difference is not statistically significant. However, for the full analysis of the parameter space, the model is mildly more unfavorable than the $\Lambda$CDM. This is evident from the fact that the probability posterior distribution for $b^2$ is inconclusive.

\subsection{Particle horizon as the length scale ($L = R_h$) }

\begin{figure}[h!]
\begin{center}
	\includegraphics[scale=0.51]{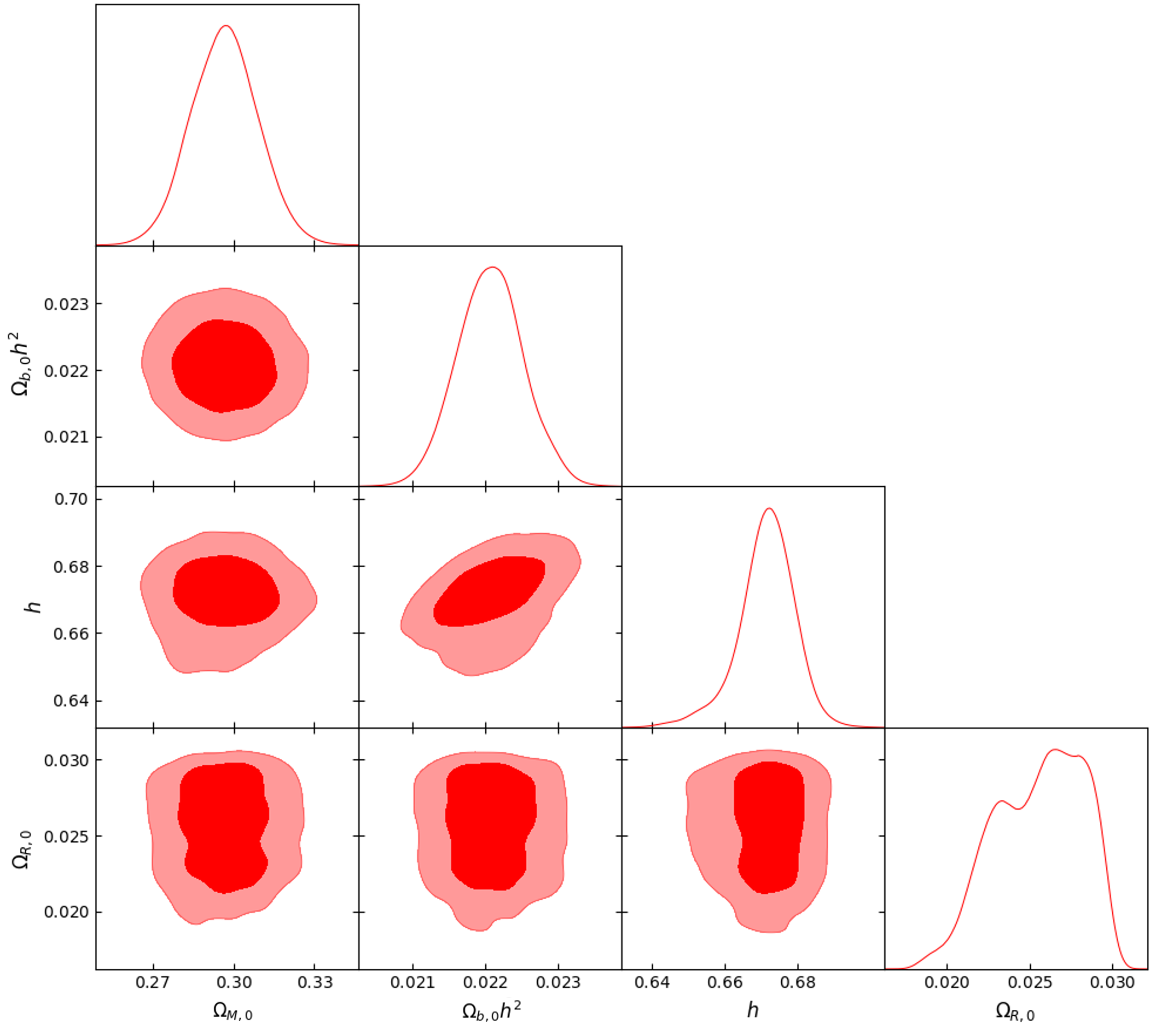}
	\includegraphics[scale=0.43]{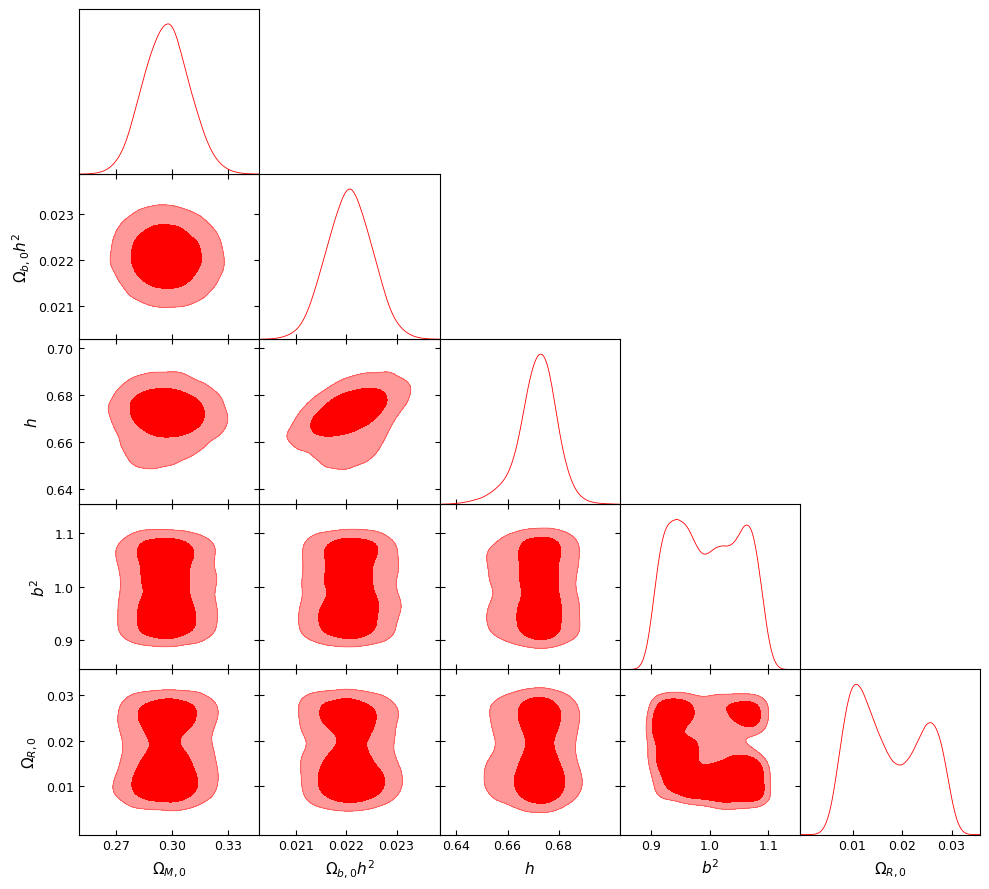}
\end{center}
\caption{Two-dimensional 68\% and 95\% C.L. allowed regions and the probability posterior distributions for model parameters from AdS-HDE: $L = R_h$ model. On the top panel, the parameter $b$ is set as unity corresponding to the discussion in Sec.~\ref{sec: partical hor}. The analysis for the full parameter space $\{\Omega_{R,0}, \Omega_{M,0}, \Omega_{b,0}, h, b^2\}$ is in the bottom panel.
}\label{Rh-contour}
\end{figure}

Unlike the previous case, $b^2$ does not rescale the density parameters defined in Sec.~\ref{sec: partical hor}. The observation values for the total density parameter for each species in this case are then straightforwardly interpreted with $\Omega_{i,0}^{\rm obs} = \Omega_{i,0}.$ The observational constraint analysis for $\text{AdS-HDE:} \;\; L = R_h$ is reported in Fig.~\ref{Rh-contour}. The central values of the allowed region of the model parameters are given in Table~\ref{Tab: model parameters-Rh} which are also consistent with the $\Lambda$CDM model. Regarding the value of AIC, as shown in Table~\ref{Tab: AIC}, the model with fixed $b^2 = 1$ is compatible with the $\Lambda$CDM. On the other hand, the full analysis shows only mildly disfavor over the $\Lambda$CDM model. Similar to the previous case, the spread of the posterior distribution of $b^2$ suggests a slightly lower statistics compared to the $\Lambda$CDM.

\begin{table}
\centering
\caption{Observational constraints on the parameters of AdS-HDE: $L = R_h$ using binned Pantheon, Cosmic chronometers, and Baryon acoustic oscillations datasets}
\begin{tabular}{|l|c|c|c|c|c|}
	\hline
Models & $\Omega_{M,0}$ & $\Omega_{b,0}h^2$ & $h$ & $b^2$ & $\Omega_{R,0}$ 
    \\\hline\hline
AdS-HDE: $L=R_h$, $b^2=1$ & $0.2955^{+0.0178}_{-0.0178}$ & $0.0221^{+0.0007}_{-0.0007}$ & $0.6700^{+0.0125}_{-0.0147}$ & 1 & $0.0257^{+0.0026}_{-0.0034}$  
    \\\hline
AdS-HDE: $L=R_h$ & \,$0.2963^{+0.0179}_{-0.0173}$\, & \,$0.0221^{+0.0006}_{-0.0006}$\, & \,$0.6704^{+0.0116}_{-0.0137}$\, & \,$1.0041^{+0.0644}_{-0.0702}$\, & \,$0.0180^{+0.0092}_{-0.0082}$\,  
    \\\hline
$\Lambda$CDM & $0.3049^{+0.0184}_{-0.0179}$ & $0.0220^{+0.0006}_{-0.0006}$ & $0.6819^{+0.0117}_{-0.0138}$ & -- & --
    \\\hline	
\end{tabular}\label{Tab: model parameters-Rh}
\end{table}
\begin{table}
\centering
\caption{Estimator of each model compared with the $\Lambda$CDM.}
\begin{tabular}{|l|c|c|}
	\hline
Models & AIC & $\Delta$AIC \\ 
 	\hline\hline
AdS-HDE: $L=H^{-1}$, $b^2=0.01$ & \,\,69.1156\,\, & \,\,$-0.5252$\,\, \\
 	\hline  
AdS-HDE: $L=H^{-1}$ & 71.6718 & 2.0310 \\
 	\hline
AdS-HDE: $L=R_h$, $b^2=1$ & 71.6244 & 1.9836 \\
 	\hline
AdS-HDE: $L=R_h$ & 73.6090 & 3.9682 \\
 	\hline
$\Lambda$CDM & 69.6408 & 0\\
 	\hline	
\end{tabular}\label{Tab: AIC}
\end{table}

\FloatBarrier

\section{Conclusion}\label{sec: conclu} 

In this work, we have proposed the HDE model by modifying the energy bound of the black hole side from the asymptotically flat black hole to the asymptotically AdS one. According to the first law of thermodynamics, it is interestingly found that there exists an additional constant term to the energy density of the original HDE model. This term actually plays an important role as a dominant content in the Universe at late time. In other words, the contribution corresponding to asymptotically dS spacetime in the cosmology viewpoint can be equivalently obtained from that of the asymptotically AdS black hole via the holographic principle. Since the cosmological length scale in the model is still arbitrary, we have chosen two simple ones which are the Hubble and the particle event horizons in order to study the dynamics of the Universe.

By choosing the Hubble horizon as the length scale, the energy density of HDE can be separated into two parts as follows: 
(i) the scaling part which tracks the evolution of the dominant content in the Universe, and
(ii) the dark energy part which drives the accelerated expansion at late time.
We also showed the exact solution of the scale factor describing the whole dynamics of the Universe from radiation dominated to dark energy dominated epochs with the suitable parameters of the model.

With the particle event horizon as the length scale, the constant term equivalent to that in the Hubble horizon case is also obtained. Therefore, the model with the particle event horizon is able to explain the late-time expansion. Another part in the expression of the energy density $\rho_R$ does not play a role in the scaling solution as seen in Eq.~\eqref{w R}. However, the non-constant contribution, $\rho_R$, in the Universe might be dominant in the period during the end of the matter dominated epoch and the beginning of the dark energy dominated epoch. Since it is not possible to solve for the exact expression of the scale factor, we have used the dynamical system approach to qualitatively analyze the dynamics of each fixed point. It was found that there are four fixed points corresponding to the domination of each content in the Universe as well as the mixture between $\Omega_M$ and $\Omega_R$. The stability of each fixed point can be compatible with the standard cosmic evolution. Even though, in the dynamical system approach, we considered the general content which could be any combination between radiation and matter, the corresponding results are in agreement with the numerical results in which the radiation and matter are treated as distinguishable contents. It is seen that the dynamics of $\Omega_R$ indeed depends on its density parameter at the present time $\Omega_{R,0}$. Moreover, the cosmic evolution of our model with the particle horizon can be reduced to that of the standard $\Lambda$CDM model by taking $\Omega_{R,0}=0$. The amount $\Omega_{R,0}$ actually modifies the amount of $\Omega_m$ in the matter dominated epoch as seen in Fig.~\ref{evo2}.

The likelihood analysis on the models parameters using the observational data has been performed. Using the Jeffreys' scale, we can conclude from Table~\ref{Tab: AIC} that the model with Hubble radius as the length scale fits the observation data as good as the $\Lambda$CDM. This result is not surprising since one can see that the act of rescaling makes this model and the $\Lambda$CDM equivalent. Although an extra number of parameters from this model is unfavorable in terms of the statistics, the model still has merit coming from the strong motivation from the holographic principle. For the model with particle horizon as the length scale, there also is no strong suggestion from the statistics that the $\Lambda$CDM performs better than this particular holographic dark energy model.

From the observation for the accelerated expansion of the Universe, the spacetime preferably corresponds to the asymptotically dS type rather than the AdS one. However, the dS counterpart of the AdS/CFT correspondence is not very well understood and has not yet been extensively studied. As a result of this work, the UV/IR relationship might reveal some connection between the AdS black hole and the dark energy associated with dS spacetime in the cosmological context. This might potentially pave the way for further investigations on the relationship between dS/CFT and AdS/CFT correspondences.

\section*{Acknowledgements}
This research project is supported by National Research Council of Thailand (NRCT) : NRCT5-RGJ63009-110. P.W. and D.S. are supported by National Science, Research and Innovation Fund (NSRF) through grant no. R2565B030. C.P. is supported by Research Grant for New Scholar, Office of the Permanent Secretary, Ministry of Higher Education, Science, Research and Innovation under contract no. RGNS 64-043. C.P. has also received funding support from the National Science, Research and Innovation Fund (NSRF).

\newpage

\nocite{*}
\bibliographystyle{mybibstyle}
\bibliography{ref.bib}

\end{document}